\documentclass[lettersize,journal]{IEEEtran}

\usepackage{cite}
\usepackage{amsmath,amssymb,amsfonts,bm}
\usepackage{mathtools}
\usepackage{multirow}
\usepackage{bigints}
\usepackage{textcomp}
\usepackage{graphicx}
\usepackage{textcomp}
\usepackage{algorithmic}
\usepackage{soul}
\usepackage{dsfont}
\usepackage{algorithm}
\usepackage{array}
\usepackage{comment}
\usepackage{esvect}
\usepackage{subcaption}
\usepackage{xcolor}
\usepackage{subcaption}
\usepackage{enumitem}
\newcolumntype{M}[1]{>{\centering\arraybackslash}m{#1}}
\usepackage{stfloats} 

\begin{document}

\title{ Degrees of Freedom of Planar \\ Holographic MIMO Systems}
%
%
%
\author{Athanasios G.~Kanatas,~\IEEEmembership{Senior Member,~IEEE}

\thanks{A. G. Kanatas is with the Department
of Digital Systems, University of Piraeus, 18534 Piraeus, Greece (e-mail:
kanatas@unipi.gr).}
}
%
%

\markboth{Journal of \LaTeX\ Class Files,~Vol.~14, No.~8, August~2021}%
{Shell \MakeLowercase{\textit{et al.}}: A Sample Article Using IEEEtran.cls for IEEE Journals}


%



\maketitle

\begin{abstract}
This paper presents a closed-form deterministic framework for scalar and vector spatial degrees of freedom (DoF) in planar holographic MIMO systems with arbitrary 3D placement and orientation. Using two continuous rectangular surfaces radiating into forward half-spaces, a geometric clipping algorithm classifies mutual visibility, extracts clipped convex apertures, and computes their effective areas and shifted centroids. A double-sided quadratic Taylor expansion anchored to these centroids reduces phase truncation and DoF leakage in near-field and heavily clipped conditions. From the continuous spatial channel kernel, we extract a geometric projection matrix whose determinant equals the cosine product of surface tilt angles relative to the effective propagation path; this quantifies orientation-induced skewing of spatial coherence cells, explains fractional DoF, and inherently avoids grazing-incidence singularities. Under partial visibility, Green’s and Parseval’s theorems show scalar DoF is shape-agnostic, scaling with effective visible area, centroid distance, and projection determinant. Extending to dual-polarized fields via the radiative dyadic Green’s function yields a Kronecker-factored tensor kernel and explicit vector DoF for geometry-dependent polarization gains.
\end{abstract}


%
\IEEEpeerreviewmaketitle

\section{Introduction}
\IEEEPARstart{H}{olographic} multiple-input multiple-output (HMIMO) has emerged as a cornerstone physical-layer paradigm for sixth-generation (6G) wireless networks \cite{Huang20}, \cite{Gong24}. Realized through electrically large, nearly continuous reconfigurable planar apertures — modeled either as continuous-aperture phased (CAP) surfaces or densely packed spatially discrete reconfigurable holographic surfaces (RHS) — HMIMO enables fine-grained synthesis and manipulation of impinging and radiated electromagnetic waves \cite{Huang20}–\cite{Deng23}. When combined with millimeter-wave (mmWave) and sub-terahertz (sub-THz) frequency bands, the large electrical dimensions of HMIMO apertures expand the radiative near-field region to tens or even hundreds of meters \cite{Liu23}, \cite{Cui23}. In this regime, spherical wavefronts replace conventional far-field plane waves, unlocking distance-dependent beam focusing and high-rank spatial multiplexing even in pure free-space line-of-sight (LoS) environments  without relying on rich multipath scattering \cite{Liu23}, \cite{Zhang23}. This is achieved by taking advantage of a significantly larger subset of the existing spatial dimensions of the aperture simultaneously \cite{Kosasih25}. Although these distinct propagation phenomena frequently create the impression that new spatial DoF emerge in the near field, this notion is a fallacy: the theoretically maximum spatial DoF of an array are governed strictly by its electrical aperture geometry and are identically achievable in both the near and far fields \cite{Kosasih25}.
\par Evaluating the fundamental performance limits of HMIMO systems falls within the domain of electromagnetic information theory (EIT), where the main metric of spatial multiplexing capability is the number of spatial DoF \cite{Marco24}, \cite{Zhu24}. Physically, DoF represent the maximum number of independent orthogonal spatial communication modes—characterized by strong coupling intensities—that can be supported concurrently between a transmitting and a receiving surface \cite{Miller00}, \cite{France18}. In rich non-line-of-sight (NLoS) scattering environments, the DoF are bounded by the electrical aperture sizes and angular scattering selectivity, scaling as $\pi S/\lambda^2$ for 2D planar apertures, where $S$ is the surface area \cite{Hu18}-\cite{Pizzo22}. In high-frequency LoS HMIMO channels, the DoF are dictated purely by deterministic wave-propagation geometry: the physical dimensions and separation distance of the surfaces, their arbitrary three-dimensional (3D) relative orientations, mutual half-space visibility constraints, and the vector polarization coupling of transverse EM fields.

Over the past two decades, several analytical and numerical methodologies have been developed to quantify the spatial DoF of holographic LoS apertures: 1) Pioneered by Miller \cite{Miller00} and expanded within EIT \cite{Marco24}, \cite{Migliore08}, the first approach casts communication between continuous regions as an integral eigenproblem governed by the free-space Green's function. While numerically solving the discretized channel operator via SVD yields exact eigenmodes and coupling strengths for arbitrary 3D placements \cite{Gong24b}, \cite{Yan23}, it incurs prohibitive computational complexity for electrically large planar surfaces and obscures explicit analytical insights into how geometric parameters govern DoF. 2) When the propagation distance significantly exceeds the aperture dimensions, the Fresnel approximation yields the classical closed-form DoF expression $S_T S_R/(\lambda d)^2$ for parallel facing surfaces of areas $S_T$ and $S_R$ separated by distance $d$ \cite{Miller00}, \cite{Miller19}. Extensions accounting for 3D array orientations under paraxial conditions \cite{Pu15}, \cite{Do23} and near-field linear or planar large intelligent surfaces links \cite{Dardari20}, \cite{Dardari21a}, provide tractable formulas, yet they typically assume fully unobstructed mutual visibility and scalar wave propagation. Moreover, expanding the propagation distance around fixed geometric aperture centers introduces severe phase truncation errors—causing beam defocusing and DoF leakage—when surfaces are heavily tilted or partially occluded at high frequencies. 3) By establishing a Fourier duality between finite spatial apertures and wavenumber-bandlimited fields, Landau's eigenvalue theorem \cite{Landau75} relates the DoF to the product of Lebesgue measures of the spectral support set and the observation set, and characterizes the asymptotic sharpness of the eigenvalue transition \cite{Landau80} - \cite{Pizzo22b}. More recently, Ruiz-Sicilia et al. \cite{Ruiz23} extended self-adjoint operator analysis to non-paraxial rotated surfaces via local partitioning. However, pre-projecting onto parallel planes absorbs 3D orientation effects implicitly, without exposing an explicit closed-form angular skewing operator, handling half-space plane clipping, or separating polarization eigenmode gains. 4) Based on the foundational works of Bucci and Franceschetti \cite{Bucci87}, \cite{Bucci89} and signal-space theory \cite{Poon05}, spatial-bandwidth methods integrate the local wavenumber support—computed via a spatially varying Jacobian determinant—across the receiving aperture \cite{Ding24} – \cite{Ruiz25}. Similarly, Gustafsson's mutual shadow-area formalism \cite{Gustafsson25} projects transmitting and receiving structures onto planes orthogonal to propagation directions to evaluate DoF for arbitrary non-paraxial and volumetric bodies. Although highly general and capable of incorporating spatial blocking \cite{Chen25}, these approaches generally require multi-dimensional numerical integrations and do not produce simple closed-form expressions linking 3D tilt angles, skewed 2D spatial coherence cells, polygonal aperture clipping, and dual-polarized vector gains. 5) In our earlier work \cite{Kanatas26}, we introduced a deterministic visibility characterization and a closed-form near-field DoF framework for 1D continuous linear arrays arbitrarily positioned and rotated on a 2D plane. However, generalizing this problem to 2D planar HMIMO surfaces in 3D space, introduces fundamentally new mathematical and physical challenges: arbitrary azimuth and zenith rotations skew the 2D diffraction lattice non-uniformly, half-space plane intersections clip rectangular apertures into arbitrary $N$-sided convex polygons whose 2D Fourier kernels are no longer separable, and planar apertures naturally support dual-polarized vector currents governed by the dyadic Green's function rather than scalar fields.
\par Despite significant progress in HMIMO channel modeling, a unified, closed-form analytical framework that simultaneously captures arbitrary 3D planar surface placements, realistic half-space radiation constraints, polygonal aperture clipping, and full dual-polarized vector EM coupling remains absent from the literature. Specifically, when two half-space radiating planar surfaces are arbitrarily translated and rotated in 3D space, the infinite plane containing one surface frequently intersects the finite boundaries of the other. Ignoring this clipping overestimates DoF, while expanding the Green's function phase around the original unclipped centers introduces truncation errors that destroy spatial orthogonality. Furthermore, the physical mechanism by which 3D tilt distorts the diffraction pattern into fractional DoF—and how transverse EM wave polarization interacts with that 3D geometry—requires an explicit, computationally lightweight mathematical characterization.
To bridge these gaps, this paper develops a comprehensive deterministic framework for evaluating the spatial DoF of the scalar and vector fields of arbitrarily positioned and oriented planar HMIMO systems in 3D space. The main novel contributions of this work are summarized as follows.
\begin{itemize}[noitemsep, topsep=0pt]
\item We formulate an unconstrained 3D geometric model for two continuous rectangular planar HMIMO surfaces radiating in single forward half-spaces, parameterized by arbitrary 3D translation coordinates and spherical azimuth/zenith rotation angles. We develop a deterministic computational geometry algorithm that evaluates normal projections and signed vertex-to-plane distances to classify full, partial, and zero visibility states, clips intersected apertures against opposing half-space cutting lines into ordered $N$-sided convex polygons, and computes their exact effective areas and updated centroids.
\item To avoid phase truncation, beam defocusing and DoF leakage during severe visibility clipping or high-frequency near-field operation, we anchor a double-sided quadratic Taylor expansion of the Euclidean path length to shifted local coordinates at the effective clipped centroids, updating the reference propagation vector and local directional cosines.
\item Using the cross-coupling phase terms of the continuous spatial channel kernel, we derive a geometric projection matrix $\mathbf{A}$ that maps physical receiver displacements to spatial frequencies. We prove that matrix $\mathbf{A}$ captures purely the alignment of projected shadows orthogonal to propagation. We demonstrate how $\mathbf{A}^{-1}$ stretches and skews the 2D lattice of kernel zeros to form skewed spatial coherence cells, physically explaining fractional degrees of freedom and weak mode coupling; while showing that the visibility algorithm inherently bypasses the singularity of grazing-incidence.
\item Under partial visibility, Green’s theorem converts the 2D diffraction integral over a clipped $N$-sided polygon into an exact sum of 1D edge integrals. Using Parseval’s theorem with the spatial-to-frequency Jacobian—and confirmed via the Hilbert–Schmidt norm and Szegö’s theorem—we show that the scalar DoF are independent of the polygon’s shape, depending only on effective areas, effective centroid distance and $\det(\mathbf{A})$.
\item We generalize the scalar analysis to full transverse vector EM fields by replacing the scalar Green's function with the radiative dyadic Green's function projected onto the local 2D surface coordinate bases. Defining the polarization coupling matrix $\mathcal{X}$ and the Gram polarization covariance matrix $\mathbf{X} = \mathcal{X}\mathcal{X}^H$, a closed-form vector DoF expression is given, which explicitly quantifies the geometry-dependent polarization gain in terms of the eigenvalues of $\mathbf{X}$.
\end{itemize}

The remainder of this paper is organized as follows. Section II models a 3D rigid-body system for two planar surfaces and derives a centroid-shifted double-sided Taylor series expansion for partial and full visibility. Section III defines the continuous spatial kernel, introduces the projection matrix $A$, derives closed-form DoF expressions for scalar fields, extends the framework to dual-polarized vector fields via the dyadic Green’s function, and provides numerical results. Section IV presents the deterministic 3D half-space visibility and polygon-clipping algorithm. Section V concludes the paper.

\section{System Model}

\begin{figure}
\centering
  \includegraphics[width=0.49\textwidth]{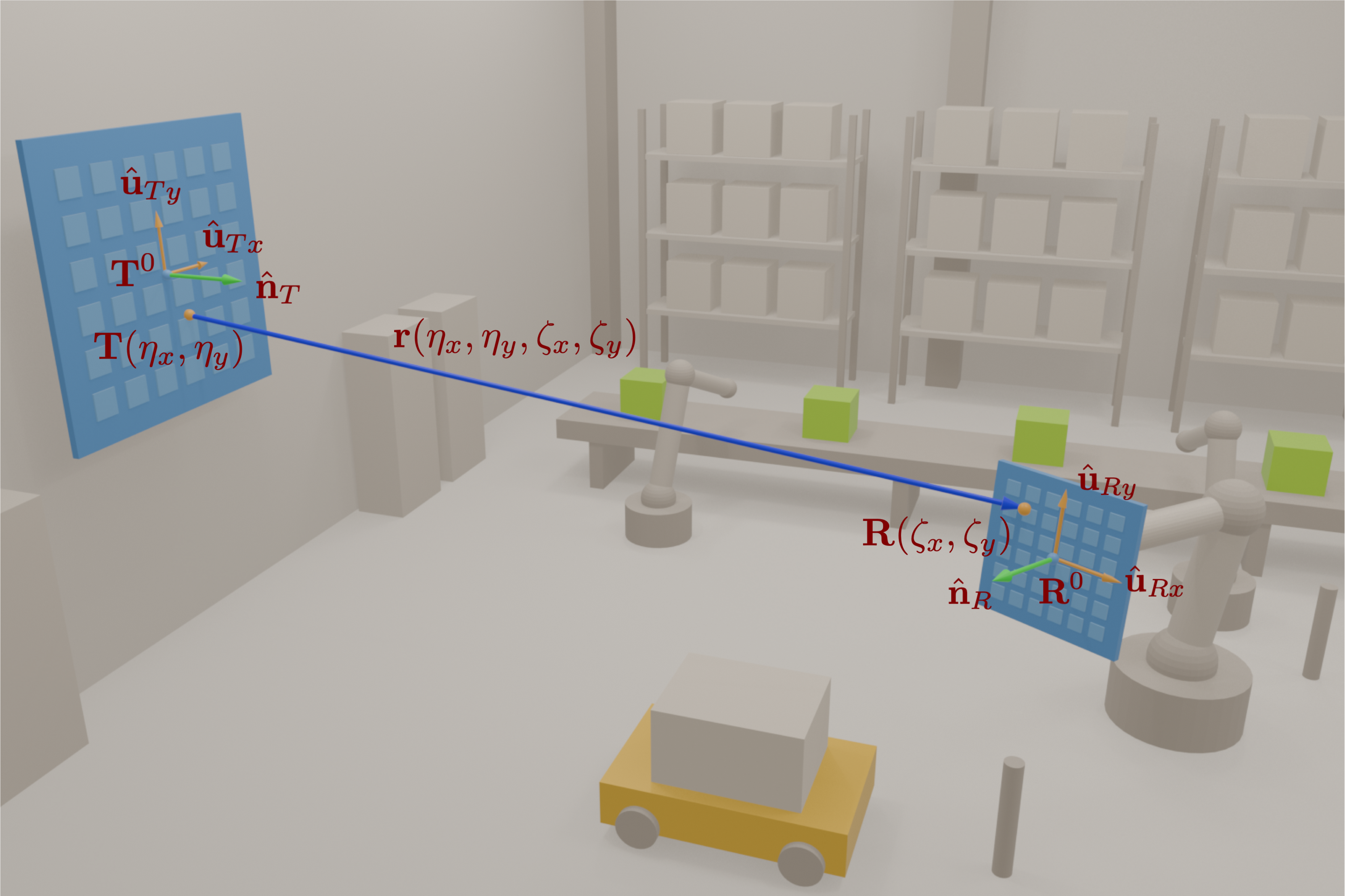}
  \caption{Considered system model depicting the capability of surfaces rotation in 3D in an industrial environment.}
  \label{fig_1}
\end{figure}

Consider two surfaces of an HMIMO system as shown in Fig. \ref{fig_1} in a 3D space defined by the fixed Cartesian unit vectors $\hat{\mathbf{x}}, \, \hat{\mathbf{y}}, \, \hat{\mathbf{z}}$. The transmitting surface is a continuous rectangular planar surface of size $L_{Tx} \times L_{Ty}$, with its center placed at the origin of the global 3D coordinate system, $\mathbf{T}^0 = (0, 0, 0)$. The orientation of the Tx plane in the 3D space is defined by two mutually orthogonal unit vectors, $\hat{\mathbf{u}}_{Tx}$ and $\hat{\mathbf{u}}_{Ty}$. The unit normal vector that determines its forward-radiating half-space is defined as the cross product of these vectors, i.e., $\hat{\mathbf{n}}_T = \hat{\mathbf{u}}_{Tx} \times \hat{\mathbf{u}}_{Ty}$. Let $\eta_x \in [-L_{Tx}/2, L_{Tx}/2]$ and $\eta_y \in [-L_{Ty}/2, L_{Ty}/2]$ denote the generic coordinates along the surface. A generic point on the Tx surface is represented by the vector: $\mathbf{T}(\eta_x, \eta_y) = \eta_x \hat{\mathbf{u}}_{Tx} + \eta_y \hat{\mathbf{u}}_{Ty}$. 
To map the local mutually orthogonal unit vectors of the transmitting surface ($\hat{\mathbf{u}}_{Tx}$ and $\hat{\mathbf{u}}_{Ty}$) to the global Cartesian unit vectors $\hat{\mathbf{x}}$, $\hat{\mathbf{y}}$, $\hat{\mathbf{z}}$, we use a 3D rotation framework parameterized by the azimuth and zenith angles of the surface. To arbitrarily orient the plane in the 3D space, we define its pointing direction (the normal vector $\hat{\mathbf{n}}_T$) using standard spherical coordinates, i.e., an azimuth angle $\phi_T$, that is, the rotation in the global $xy$-plane measured from the $x$-axis, and a zenith angle $\theta_T$, which is the angle measured downward from the global $z$-axis. Moreover, to ensure that the local basis vectors span the plane perfectly orthogonal to $\hat{\mathbf{n}}_T$, they are aligned with the standard spherical coordinate unit vectors $\hat{\pmb{\theta}}$ and $\hat{\pmb{\phi}}$.
In this way, the local $x$-axis of the surface ($\hat{\mathbf{u}}_{Tx}$) is assigned to the "vertical/zenith" tilt, and the local $y$-axis of the surface ($\hat{\mathbf{u}}_{Ty}$) to the "horizontal/azimuthal" sweep. The matrix that rotates a surface originally lying flat in the global $xy$-plane into its final skewed 3D orientation is given by 
\begin{equation}\label{rotation}
\begin{bmatrix} \hat{\mathbf{u}}_{Tx} \\ \hat{\mathbf{u}}_{Ty} \\ \hat{\mathbf{n}}_T \end{bmatrix} = \begin{bmatrix} \cos\theta_T \cos\phi_T & \cos\theta_T \sin\phi_T & -\sin\theta_T \\ -\sin\phi_T & \cos\phi_T & 0 \\ \sin\theta_T \cos\phi_T & \sin\theta_T \sin\phi_T & \cos\theta_T \end{bmatrix} \begin{bmatrix} \hat{\mathbf{x}} \\ \hat{\mathbf{y}} \\ \hat{\mathbf{z}} \end{bmatrix}	.
\end{equation}

\par The receiving surface is a continuous rectangular planar surface of size $L_{Rx} \times L_{Ry}$. Its center is located at the Cartesian coordinates $(x_0, y_0, z_0)$, which defines the vector $\mathbf{R}^0= x_0\hat{\mathbf{x}} + y_0\hat{\mathbf{y}} + z_0\hat{\mathbf{z}}$. The orientation of the Rx plane is defined by two mutually orthogonal unit vectors, $\hat{\mathbf{u}}_{Rx}$ and $\hat{\mathbf{u}}_{Ry}$. Its unit normal vector is $\hat{\mathbf{n}}_R = \hat{\mathbf{u}}_{Rx} \times \hat{\mathbf{u}}_{Ry}$. Let $\zeta_x \in [-L_{Rx}/2, L_{Rx}/2]$ and $\zeta_y \in [-L_{Ry}/2, L_{Ry}/2]$ denote the generic coordinates along the receiving surface. A generic point on the Rx surface is represented by the vector $\mathbf{R}(\zeta_x, \zeta_y) = \mathbf{R}^0 + \zeta_x \hat{\mathbf{u}}_{Rx} + \zeta_y \hat{\mathbf{u}}_{Ry}$. The exact same angular parameterization applies to the receiving surface. In 3D space, orienting and positioning the receiver requires a rigid-body transformation, i.e., a rotation to define its local vectors using its own azimuth and zenith angles followed by a translation to place its center at the specific spatial coordinates. Using an identical spherical-to-Cartesian mapping, the local mutually orthogonal unit vectors $\hat{\mathbf{n}}_R$,  $\hat{\mathbf{u}}_{Rx}$ and $\hat{\mathbf{u}}_{Ry}$, for the Rx plane, are given by a matrix similar to that in (\ref{rotation}) by substituting the angles $\theta_T$ and $\phi_T$ with $\theta_R$ and $\phi_R$ correspondingly. It is noted that $\hat{\mathbf{u}}_{Ry}$ is explicitly forced to have zero $\hat{\mathbf{z}}$-component, and this means that the local $y$-axis always lies perfectly parallel to the global horizontal $xy$-plane, regardless of how the surface is tilted vertically. It is determined entirely by rotating away from the global $\hat{\mathbf{y}}$-axis by the azimuth angle $\phi_R$. Because the local $y$-axis is locked horizontally, the local $x$-axis absorbs all the vertical tilt. Thus, the angle $\theta_R$ dictates how much this local axis is tilted up or down relative to the global horizontal plane.
\par The Euclidean distance connecting a generic point on the transmitting surface to a generic point on the receiving surface is given by the norm of the distance vector:
\begin{equation} \label{Distance_1}
	\begin{aligned}
	\mathbf{r}(\eta_x, \eta_y, \zeta_x, \zeta_y) &= \mathbf{R}(\zeta_x, \zeta_y) - \mathbf{T}(\eta_x, \eta_y) \\
	& =\mathbf{R}^0 + \zeta_x \hat{\mathbf{u}}_{Rx} + \zeta_y \hat{\mathbf{u}}_{Ry} - \eta_x \hat{\mathbf{u}}_{Tx} - \eta_y \hat{\mathbf{u}}_{Ty}.
	\end{aligned}
\end{equation}
Using the $x,y,z$-components of the local unit vectors
\begin{equation} \label{Norm_Distance_1}
\begin{aligned}
	&\vert{}\vert{}\mathbf{r}(\eta_x, \eta_y, \zeta_x, \zeta_y)\vert{}\vert{} =\\ &\sqrt{ \begin{aligned} & (x_0 + \zeta_x u_{Rx,x} + \zeta_y u_{Ry,x} - \eta_x u_{Tx,x} - \eta_y u_{Ty,x})^2 \\ + & (y_0 + \zeta_x u_{Rx,y} + \zeta_y u_{Ry,y} - \eta_x u_{Tx,y} - \eta_y u_{Ty,y})^2 \\ + & (z_0 + \zeta_x u_{Rx,z} + \zeta_y u_{Ry,z} - \eta_x u_{Tx,z} - \eta_y u_{Ty,z})^2 \end{aligned} }.
	\end{aligned}
\end{equation}
\begin{figure*}[!ht]
	\centering
	\begin{subfigure}[t]{.4\linewidth}
		\includegraphics[trim=0 0 0 0,clip,width=\linewidth]{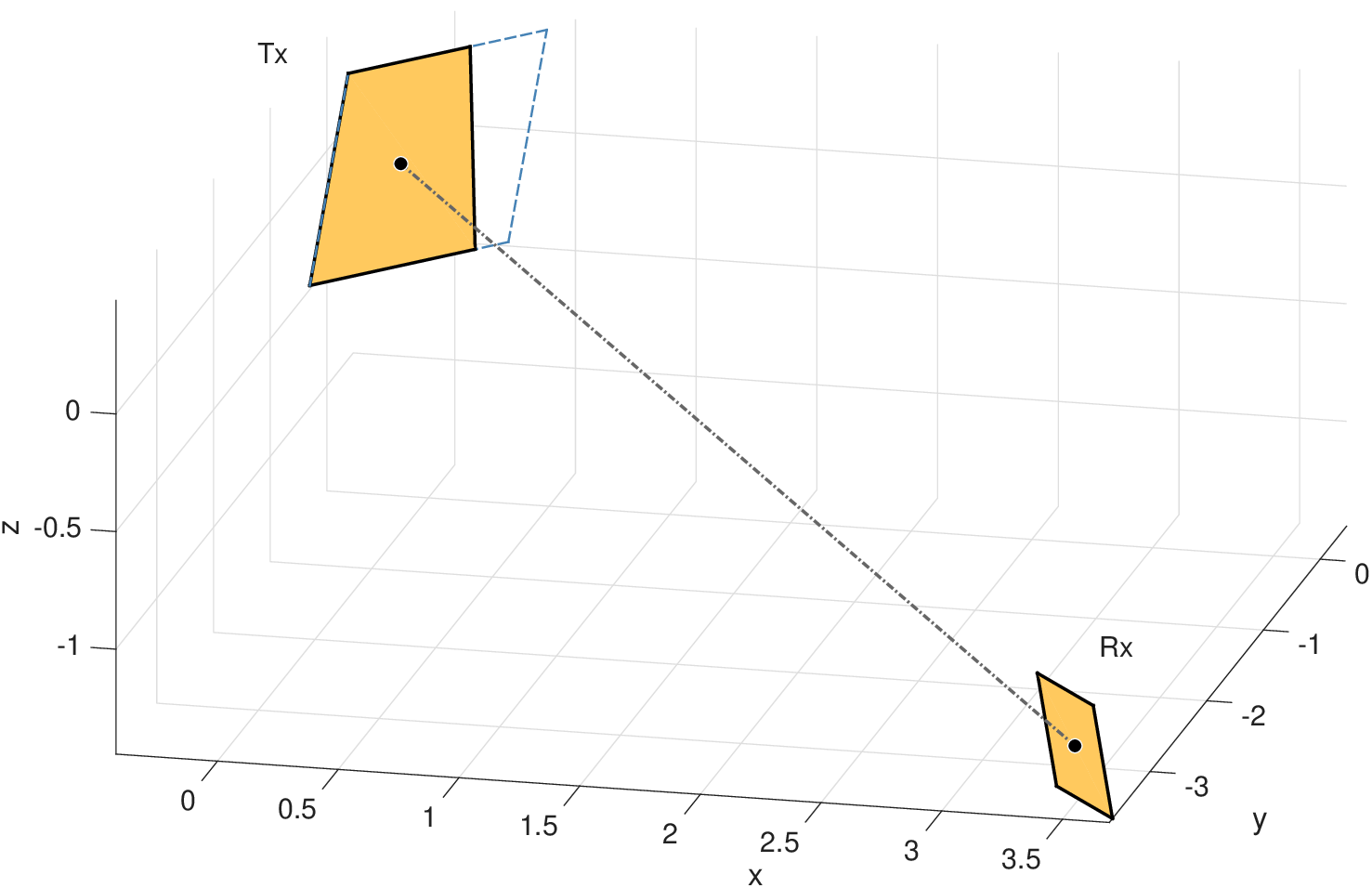}
		\caption{The Tx is intersected}
		\label{fig_2_a}
	\end{subfigure}\hfil
	\begin{subfigure}[t]{.4\linewidth}
		\includegraphics[trim=0 0 0 0,clip,width=\linewidth]{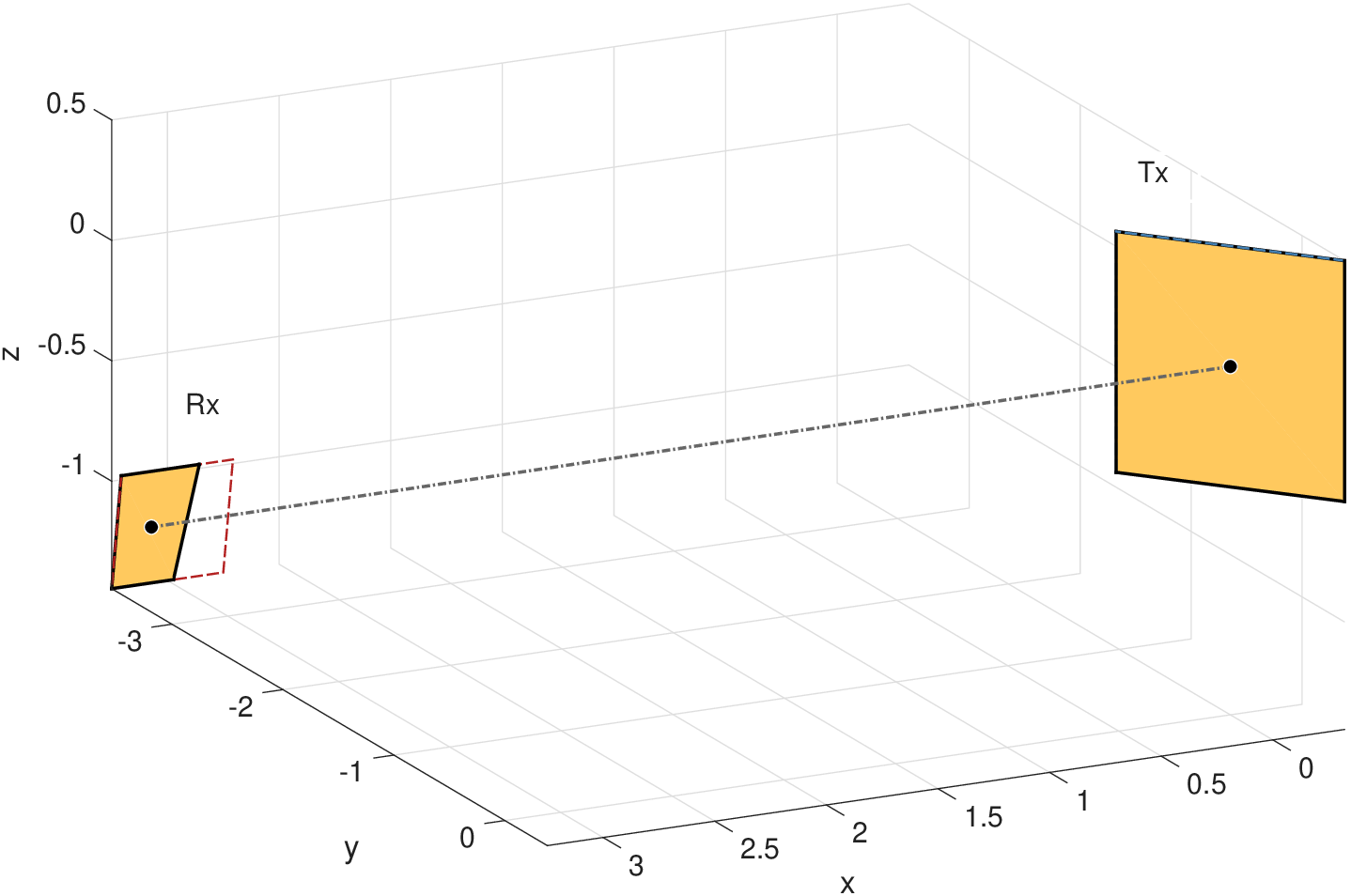}%
		\caption{The Rx is intersected}
		\label{fig_2_b}
	\end{subfigure}
	\caption{Two case studies with clipped effective regions are depicted, $L_{Tx}=1m$, $L_{Ty}=1m$, $L_{Rx}=0.5m$, $L_{Ry}=0.5m$ . a) $\theta_T=105^o$, $\phi_T=-40^o$,$\theta_R=100^o$, $\phi_R=220^o$, $x_0=3.5$, $y_0=-3.5$, $z_0=-1.2$, b) $\theta_T=90^o$, $\phi_T=40^o$,$\theta_R=100^o$, $\phi_R=90^o$, $x_0=3.0$, $y_0=-3.5$, $z_0=-1.2$.}
	\label{Intersections}
\end{figure*}
In the presented model, the receiving surface can be located anywhere in the 3D space, and the two surfaces may be rotated in azimuth and/or elevation. Consequently, it is possible that the receiving side of one surface may not be fully visible from the transmitting side of the other, and vice versa. Typical examples are depicted in Fig.\ref{Intersections}, where the visible regions are depicted for two cases. Interestingly, based on the values of $x_0, y_0, z_0, \theta_T, \theta_R, \phi_T, \phi_R$, three states are possible: full, partial, or no visibility, and thus it is necessary to calculate the visible (effective) areas of the two arrays, $S_{T,eff}$, $S_{R,eff}$, i.e., the clipped geometric polygon areas on the surfaces that constitute the effective transmitting and receiving regions. In the case of partial visibility, the new effective centers $\mathbf{T}^0_{eff}$, $\mathbf{R}^0_{eff}$, which are the geometric centroids of the clipped surfaces, should also be calculated.
\par A typical focusing function towards a point $ \mathbf{R}(\zeta_x, \zeta_y)$ on the receiving surface is given by
\begin{equation}
	F_{Tx}(\eta_x, \eta_y)\vert{}_\mathbf{R} = \frac{1}{\sqrt{S_{T,eff}}}e^{jk\vert{}\vert{}\mathbf{r}(\eta_x,\eta_y,\zeta_x,\zeta_y)\vert{}\vert{}}.
\end{equation}
Analytically evaluating this focusing operation and the resulting spatial interaction between the HMIMO surfaces requires approximating the Euclidean distance $\|\mathbf{r}(\eta_x,\eta_y,\zeta_x,\zeta_y)\|$ via a double-sided quadratic Taylor series expansion. If visibility is full or high -- meaning that the effective centroid $\mathbf{T}_{eff}^0$ remains close to $\mathbf{T}^0$ -- or if the Rx surface lies in the far field of the Tx surface, expanding the distance around the original center $(\eta_x=0,\eta_y=0)$ remains accurate. Conversely, when visibility is severely clipped in the near field, expanding around $(0,0)$ leaves the active region $S_{T,eff}$ offset from the expansion point, amplifying the omitted higher-order Taylor truncation terms. Consequently, $F_{Tx}(\eta_x,\eta_y)|_R$ fails to accurately match the spherical wavefront curvature across the clipped aperture, causing beam defocusing and DoF leakage as phase-mismatches destroy the orthogonality of the received phase signatures. Then, the expansion should be made around the shifted local coordinates $(\tilde{\eta}_x=0,\tilde{\eta}_y=0)$ defined by
\begin{equation}\label{shiftvartx}
\tilde{\eta}_x = \eta_x - \eta_{x,c} \quad \text{and} \quad \tilde{\eta}_y = \eta_y - \eta_{y,c},
\end{equation}
where $(\eta_{x,c},\eta_{y,c})$ are the local coordinates of $\mathbf{T}_{eff}^0$.
\subsection{Partial Visibility Taylor Series Expansion}
To expand the distance function around the new effective centers under partial visibility, we define a shifted local coordinate system anchored to the centroids of the clipped polygons. Let the geometric centroids of the clipped Tx and Rx surfaces in their original local coordinates be $(\eta_{x,c}, \eta_{y,c})$ and $(\zeta_{x,c}, \zeta_{y,c})$. The shifted local variables at the Tx are given by (\ref{shiftvartx}), whereas at the Rx are given by
\begin{equation}\label{shiftvarrx}
	\tilde{\zeta}_x = \zeta_x - \zeta_{x,c} \,\,\,\,\rm{and} \,\,\,\,  \tilde{\zeta}_y = \zeta_y - \zeta_{y,c}
\end{equation}
where $(\zeta_{x,c},\zeta_{y,c})$ are the local coordinates of $\mathbf{R}^0_{eff}$ for which the new global coordinates are given by
\begin{equation}\label{eff0rx}
	\mathbf{R}^0_{eff}=\mathbf{R}^0 + \zeta_{x,c}\hat{\mathbf{u}}_{Rx} + \zeta_{y,c}\hat{\mathbf{u}}_{Ry}.
\end{equation}
The reference distance vector that connects the new centroids on the two surfaces is given by
\begin{equation}\label{refdist}
	\mathbf{r}_{eff}=\mathbf{R}^0_{eff}-\mathbf{T}^0_{eff}
\end{equation}
with $d_{eff}=\vert{}\vert{}\mathbf{r}_{eff}\vert{}\vert{}$ and the corresponding unit vector given by $\hat{\mathbf{r}}_{eff} = \mathbf{r}_{eff}/d_{eff}$.
Any point on the Tx and Rx surfaces can now be expressed using the shifted variables
\begin{equation}
	\begin{aligned}
	&\mathbf{T}(\tilde{\eta}_x, \tilde{\eta}_y) = \mathbf{T}^0_{eff} + \tilde{\eta}_x \hat{\mathbf{u}}_{Tx} + \tilde{\eta}_y \hat{\mathbf{u}}_{Ty} \\
	&\mathbf{R}(\tilde{\zeta}_x, \tilde{\zeta}_y) = \mathbf{R}^0_{eff} + \tilde{\zeta}_x \hat{\mathbf{u}}_{Rx} + \tilde{\zeta}_y \hat{\mathbf{u}}_{Ry}	
	\end{aligned}.
\end{equation}
The vector between these points is 
\begin{equation}
	\begin{aligned}
	\mathbf{r}(\tilde{\eta}_x, \tilde{\eta}_y,\tilde{\zeta}_x, \tilde{\zeta}_y)&=\mathbf{r}_{eff} + \tilde{\zeta}_x \hat{\mathbf{u}}_{Rx} + \tilde{\zeta}_y \hat{\mathbf{u}}_{Ry} - \tilde{\eta}_x \hat{\mathbf{u}}_{Tx} - \tilde{\eta}_y \hat{\mathbf{u}}_{Ty} \\
	& = \mathbf{r}_{eff} + \pmb{\Delta}
		\end{aligned}
\end{equation}
where $\pmb{\Delta}$ is the spatial displacement vector. By applying the binomial expansion, it is easily shown that for $\Vert{}\mathbf{\Delta}\Vert{} \ll d_{eff}$ and keeping the first three terms, i.e., up to the quadratic term,
\begin{equation}\label{Taylor}
\Vert{}\mathbf{r}_{eff} + \mathbf{\Delta}\Vert{} \approx d_{eff} + \hat{\mathbf{r}}_{eff} \cdot \mathbf{\Delta} + \frac{1}{2 d_{eff}} \left( \Vert{}\mathbf{\Delta}\Vert{}^2 - (\hat{\mathbf{r}}_{eff} \cdot \mathbf{\Delta})^2 \right).
\end{equation}
It is easily observed that the Euclidean distance between the two points on the two surfaces is broken into three components, a constant, a linear far-field phase projection, and a quadratic term that represents the near-field wavefront curvature correction. The linear term is expanded as follows
\begin{equation}\label{dlinear}
	d_{linear}=\hat{\mathbf{r}}_{eff} \cdot \mathbf{\Delta}=  \tilde{\rho}_{Rx}\tilde{\zeta}_x + \tilde{\rho}_{Ry}\tilde{\zeta}_y -\tilde{\rho}_{Tx}\tilde{\eta}_x - \tilde{\rho}_{Ty}\tilde{\eta}_y
\end{equation}
where the terms $\tilde{\rho}_{Tx}, \tilde{\rho}_{Ty}, \tilde{\rho}_{Rx}, \tilde{\rho}_{Ry}$ are the directional cosines that define how the reference propagation path ($\hat{\mathbf{r}}_{eff}$) projects onto the local axes of the transmitting and receiving surfaces, and are given by
\begin{equation}
	\begin{aligned}
	&\tilde{\rho}_{Ti} = \hat{\mathbf{r}}_{eff} \cdot \hat{\mathbf{u}}_{Ti} \,\,\,{\rm{for}} \,\,\, i \in\{x,y\}\\
	&\tilde{\rho}_{Rj} = \hat{\mathbf{r}}_{eff} \cdot \hat{\mathbf{u}}_{Rj} \,\,\,{\rm{for}} \,\,\, j \in\{x,y\}.
	\end{aligned}
\end{equation}
Similarly, the quadratic term is expanded as
\begin{equation}\label{dquad}
	\begin{aligned}
	& d_{quad} =  \frac{1}{2 d_{eff}} \left( \Vert{}\mathbf{\Delta}\Vert{}^2 - (\hat{\mathbf{r}}_{eff} \cdot \mathbf{\Delta})^2 \right) \\
	 = &\frac{1}{2d_{eff}} \Bigg\{ \Big[ \tilde{\eta}_x^2(1 - \tilde{\rho}_{Tx}^2) + \tilde{\eta}_y^2(1 - \tilde{\rho}_{Ty}^2) - 2\tilde{\eta}_x\tilde{\eta}_y(\tilde{\rho}_{Tx}\tilde{\rho}_{Ty}) \Big] \\
	& + \Big[ \tilde{\zeta}_x^2(1 - \tilde{\rho}_{Rx}^2) + \tilde{\zeta}_y^2(1 - \tilde{\rho}_{Ry}^2) - 2\tilde{\zeta}_x\tilde{\zeta}_y(\tilde{\rho}_{Rx}\tilde{\rho}_{Ry}) \Big] \\
	& - 2 \sum_{i \in \{x,y\}} \sum_{j \in \{x,y\}} \tilde{\eta}_i \tilde{\zeta}_j \Big( \hat{\mathbf{u}}_{Ti} \cdot \hat{\mathbf{u}}_{Rj} - \tilde{\rho}_{Ti}\tilde{\rho}_{Rj} \Big) \Bigg\}
	\end{aligned}.
\end{equation}
\subsection{Full Visibility Taylor Series Expansion}
When a full visibility condition applies, then the shifted local variables revert to the physical local variables $\tilde{\eta}_x=\eta_x$, $\tilde{\eta}_y=\eta_y$, $\tilde{\zeta}_x=\zeta_x$, $\tilde{\zeta}_y=\zeta_y$, the effective reference centers revert to the original global centers, $\mathbf{T}^0_{eff}=\mathbf{T}^0$, $\mathbf{R}^0_{eff}=\mathbf{R}^0$, the reference distance vector becomes $\mathbf{r}_{eff}=\mathbf{r}_0= \mathbf{R}^0$ and $d_0=\vert{}\vert{}\mathbf{r}_0\vert{}\vert{}$, and the Taylor series expansion of the distance is given by (\ref{Taylor}) with (\ref{dlinear}) and (\ref{dquad}) evaluated under the substitution $\tilde{\rho} \rightarrow \rho$.
\section{Calculation of the DoF}
To evaluate the DoF for arbitrarily placed HMIMO planar surfaces, the communication channel is modeled as a continuous spatial operator governed by the free-space Green's function. The methodology follows earlier research work by \cite{Miller00}, \cite{Dardari20}, and \cite{Kanatas26}. The electromagnetic field at any point on the receiving surface is formed by the continuous superposition of waves radiating from the transmitting area. By evaluating the spatial correlation between any two generic observation points on the receiving plane, we formulate a continuous spatial kernel function. This kernel is mathematically defined as the surface integral of the conjugate product of the Green's functions over the effective transmitting area, capturing the complete spatial covariance of the channel.
In the following, it is shown that by substituting the double-sided distance expansion into the Green's function phase, the kernel integral resolves into a 2D spatial diffraction pattern. The spatial DoF is then extracted as the rank of this continuous kernel operator. By mapping the kernel over the effective receiving geometry to form a spatial covariance matrix, the DoF is calculated by performing an eigenvalue decomposition and counting the non-negligible eigenmodes, which correspond to the maximum number of independent, orthogonal spatial channels supported by the physical zero-crossings of the interference pattern.
\subsection{Scalar Electromagnetic Fields}
The free space Green's function for a generic point $\mathbf{T}(\tilde{\eta}_x,\tilde{\eta}_y)$ on the Tx surface and a point $\mathbf{R}(\tilde{\zeta}_x,\tilde{\zeta}_y)$ on the Rx surface is given by
\begin{equation}
	G(\tilde{\pmb{\eta}},\tilde{\pmb{\zeta}})=\frac{\rm{exp}(-jk\vert{}\vert{}\mathbf{r}(\tilde{\eta}_x, \tilde{\eta}_y,\tilde{\zeta}_x, \tilde{\zeta}_y)\vert{}\vert{})}{4\pi \vert{}\vert{}\mathbf{r}(\tilde{\eta}_x, \tilde{\eta}_y,\tilde{\zeta}_x, \tilde{\zeta}_y)\vert{}\vert{}}
\end{equation}
and the channel kernel function for two points $\tilde{\pmb{\zeta}}=(\tilde{\zeta}_x,\tilde{\zeta}_y),\tilde{\pmb{\zeta}}'=(\tilde{\zeta}'_x, \tilde{\zeta}'_y)$ on the Rx surface is given by
\begin{equation} \label{kernel_1}
K(\tilde{\pmb{\zeta}},\tilde{\pmb{\zeta}}') = \iint_{S_{T,eff}} \frac{{\rm{exp}}\left[-jk \left( d(\tilde{\pmb{\eta}},\tilde{\pmb{\zeta}}) - d(\tilde{\pmb{\eta}},\tilde{\pmb{\zeta}}') \right)\right]}{(4\pi)^2 d(\tilde{\pmb{\eta}},\tilde{\pmb{\zeta}})  d(\tilde{\pmb{\eta}},\tilde{\pmb{\zeta}}')} \rm{d}\tilde{\eta}_x \rm{d}\tilde{\eta}_y
\end{equation}
where $d(\tilde{\pmb{\eta}},\tilde{\pmb{\zeta}}) = \vert{}\vert{}\mathbf{r}(\tilde{\eta}_x, \tilde{\eta}_y,\tilde{\zeta}_x, \tilde{\zeta}_y)\vert{}\vert{}$ and $d(\tilde{\pmb{\eta}},\tilde{\pmb{\zeta}}') = \vert{}\vert{}\mathbf{r}(\tilde{\eta}_x, \tilde{\eta}_y,\tilde{\zeta}'_x, \tilde{\zeta}'_y)\vert{}\vert{}$. Since $\Vert{}\mathbf{\Delta}\Vert{} \ll d_{eff}$, $\Vert{}\mathbf{\Delta}'\Vert{} \ll d_{eff}$, the distances in the denominators can be approximated by the distance between the centroids of the two surfaces, i.e., $d(\tilde{\pmb{\eta}},\tilde{\pmb{\zeta}}) \approx d(\tilde{\pmb{\eta}},\tilde{\pmb{\zeta}}') \approx d_{eff}$. As explained in \cite{Sherman62}, if the distance $d_{eff}$ satisfies $d_{eff} \geq 1.2 \, (D_T + D_R)$, where $D_T,D_R$ are the maximum linear dimensions of the two surfaces, the variations of the amplitude are negligible and thus the approximation for the amplitude is valid. This approximation cannot be adopted for the phases in (\ref{kernel_1}).
The constant scalar $1/\left(4 \pi d_{eff}\right)^2$ represents the baseline path loss between the visible centers. Because scalar multipliers do not change the rank of a matrix, the DoF are dictated entirely by the spatial diffraction pattern generated by the remaining exponential phase terms integrated over the clipped area. In Appendix A it is shown that the magnitude of the kernel function can be written as follows:
\begin{equation} \label{kernel_2}
\vert{}K(\Delta\tilde{\zeta}_x,\Delta\tilde{\zeta}_y)\vert{} \propto \left\vert{} \iint_{S_{T,eff}} e^{j (\tilde{K}_{\rho,x}\tilde{\eta}_x + \tilde{K}_{\rho,y}\tilde{\eta}_y)} d\tilde{\eta}_x d\tilde{\eta}_y \right\vert{}
\end{equation}
where $\Delta\tilde{\zeta}_x = \tilde{\zeta}_x -\tilde{\zeta}'_x$, $\Delta\tilde{\zeta}_y = \tilde{\zeta}_y -\tilde{\zeta}'_y$, and the spatial frequencies are given by
\begin{equation}\label{Krhox}
	\begin{aligned}
	\tilde{K}_{\rho,x} = \frac{k}{d_{eff}} & \Bigl[ \Delta\tilde{\zeta}_x \left(\hat{\mathbf{u}}_{Tx} \cdot \hat{\mathbf{u}}_{Rx} - \tilde{\rho}_{Tx}\tilde{\rho}_{Rx}\right) + \bigr. \\ 
	& \bigl. \Delta\tilde{\zeta}_y \left(\hat{\mathbf{u}}_{Tx} \cdot \hat{\mathbf{u}}_{Ry} - \tilde{\rho}_{Tx}\tilde{\rho}_{Ry}\right) \Bigr] \\
	\end{aligned}
\end{equation}
\begin{equation}\label{Krhoy}
	\begin{aligned}
	\tilde{K}_{\rho,y} = \frac{k}{d_{eff}} \Bigl[ & \Delta\tilde{\zeta}_x (\hat{\mathbf{u}}_{Ty} \cdot \hat{\mathbf{u}}_{Rx} - \tilde{\rho}_{Ty}\tilde{\rho}_{Rx}) + \bigr. \\ 
	 &\bigl. \Delta\tilde{\zeta}_y (\hat{\mathbf{u}}_{Ty} \cdot \hat{\mathbf{u}}_{Ry} - \tilde{\rho}_{Ty}\tilde{\rho}_{Ry}) \Bigr]
	 \end{aligned}.	
\end{equation}
In order to depict the effect of the relative geometry of the two surfaces on the spatial frequencies and the DoF we construct a projection matrix $\mathbf{A}$ whose elements are the following 
\begin{equation}
	A_{ij}= \hat{\mathbf{u}}_{Ti} \cdot \hat{\mathbf{u}}_{Rj} - \tilde{\rho}_{Ti}\tilde{\rho}_{Rj}, \,\,\, i,j \in \{x,y\}
\end{equation}
The dot product captures the 2D twist or roll between the two surfaces and measures how well the local axes physically align in the 3D space. The second term measures the surface tilt relative to the unit vector $\mathbf{\hat{r}}_{eff}$ along the centroid-connecting vector. 
The spatial frequency vector can now be written as
\begin{equation}
	\mathbf{\tilde{K}_{\rho}} = \begin{bmatrix} \tilde{{K}}_{\rho,x} \\ \tilde{{K}}_{\rho,y} \end{bmatrix} = \frac{k}{d_{eff}} \mathbf{A}\begin{bmatrix} \Delta\tilde{\zeta}_x \\ \Delta\tilde{\zeta}_y \end{bmatrix} = \frac{k}{d_{eff}} \mathbf{A} \Delta\tilde{\pmb{\zeta}}.
\end{equation}
\subsubsection{DoF in Full Visibility}
When full visibility is obtained, the integration domain is the initial physical rectangular boundaries of the Tx surface. Spatial frequencies $K_{\rho,x}$ and $K_{\rho,y}$ are calculated using the original unshifted directional cosines ($\rho$). Because the boundaries are symmetric and orthogonal, the 2D integral separates perfectly into two independent integrals, and the magnitude of the kernel evaluates to the product of two $\rm{sinc}$ functions, where ${\rm{sinc}(x)}\triangleq\frac{\sin (\pi x)}{\pi x}$:
\begin{equation}\label{Full_Kernel}
\vert{}K_{full}(\Delta\zeta_x, \Delta\zeta_y)\vert{} \propto \left\vert{} \text{sinc}\left( \frac{K_{\rho,x} L_{Tx}}{2\pi} \right) \text{sinc}\left( \frac{K_{\rho,y} L_{Ty}}{2\pi} \right) \right\vert{}.
\end{equation}
The zeros of this kernel form a perfect rectangular lattice. The number of DoF is strictly determined by how many of these discrete nulls physically fit inside the receiver's unclipped area $S_R=L_{Rx} \times L_{Ry}$. To analytically calculate the spatial zeros of the kernel one should evaluate the conditions where the magnitude of the dual ${\rm{sinc}}$ function drops to zero, i.e.,
\begin{equation}
	\frac{K_{\rho,x} L_{Tx}}{2} = n_x \pi \quad \text{and} \quad \frac{K_{\rho,y} L_{Ty}}{2} = n_y \pi
\end{equation}
where $n_x, n_y \in \{\pm 1, \pm 2, \pm 3, \dots\}$. We can express the zero-crossings on the Rx surface as a system of linear equations:
\begin{equation}
	\begin{aligned}
& \begin{bmatrix} A_{xx} & A_{xy} \\ A_{yx} & A_{yy} \end{bmatrix} \begin{bmatrix} \Delta\zeta_x \\ \Delta\zeta_y \end{bmatrix} = \lambda d_{0} \begin{bmatrix} \frac{n_x}{L_{Tx}} \\ \frac{n_y}{L_{Ty}} \end{bmatrix}	\\ & \Rightarrow \begin{bmatrix} \Delta\zeta_x \\ \Delta\zeta_y \end{bmatrix} = \mathbf{A}^{-1} \begin{bmatrix} n_x \frac{\lambda d_0}{L_{Tx}} \\ n_y \frac{\lambda d_0}{L_{Ty}} \end{bmatrix}
	\end{aligned}.
\end{equation}
Because $n_x$ and $n_y$ are independent integers, this equation defines a skewed, regular 2D lattice on the receiving surface. Defining a spatial coherence grid cell as the area between adjacent zeros in both dimensions, ${\rm{det}}\left(\mathbf{A}^{-1}\right)$ scales the area of each cell, directly accounting for the angular foreshortening of the misaligned surfaces. Therefore, when the two surfaces are perfectly aligned, i.e., they are parallel and facing each other directly on-axis, the normal vectors align with the propagation path, so $\rho_{Tx} = \rho_{Ty} = \rho_{Rx} = \rho_{Ry} = 0$, the local axes perfectly align, and $\mathbf{A}={\rm{diag}}(1,-1)$. The zero locations then simplify to:
\begin{equation}
\Delta\zeta_x = n_x \frac{\lambda d_0}{L_{Tx}}, \quad \text{and} \quad
\Delta\zeta_y = n_y \frac{\lambda d_0}{L_{Ty}}	.
\end{equation}
The physical dimensions of a single spatial coherence cell, which is the physical area required to support one independent spatial mode, is simply :
\begin{equation}
A_{coh} = \frac{(\lambda d_0)^2}{L_{Tx} L_{Ty}} = \frac{(\lambda d_0)^2}{S_{T}}.
\end{equation}
Thus, DoF is strictly the number of these independent coherence cells that can tile across the physical area of the receiving surface ($S_{R} = L_{Rx} \times L_{Ry}$) and can be calculated by:
\begin{equation}
\text{DoF}_{scalar} \approx \frac{S_{R}}{A_{coh}} = \frac{S_{R}}{\frac{(\lambda d_0)^2}{S_{T}}} = \frac{S_{T} S_{R}}{(\lambda d_0)^2}.
\end{equation}
This recovers the theoretical scalar spatial multiplexing limit directly from the diffraction zeros of the kernel.
\par When surfaces are misaligned, the projection matrix $\mathbf{A}$ skews and stretches the 2D zero grid. The area of the skewed coherence cell is modified as follows:
\begin{equation}\label{Acohskewed}
	A_{coh, skewed} = \left|\det(\mathbf{A}^{-1})\right| \frac{(\lambda d_0)^2}{S_{T}} = \frac{1}{\left|\det(\mathbf{A})\right|} \frac{(\lambda d_0)^2}{S_{T}}.
\end{equation}
Consequently, the DoF is calculated by the following:
\begin{equation}\label{DOFscalar_full}
	\text{DoF}_{scalar} = \left|\det(\mathbf{A})\right| \frac{S_{T} S_{R}}{(\lambda d_0)^2}.
\end{equation}
In Appendix B, it is proved that
\begin{equation}\label{detA}
	|{\rm{det}}\mathbf{A}| = |\left(\hat{\mathbf{n}}_T \cdot \hat{\mathbf{r}}_{eff} \right)\left(\hat{\mathbf{n}}_R \cdot \hat{\mathbf{r}}_{eff}\right)| = \cos(\gamma_T) \cos(\gamma_R)
\end{equation}
where the tilt angles $\gamma_T$ and $\gamma_R$ are the geometric 3D angles that quantify the tilt from the effective line-of-sight path, and $\gamma_T, \gamma_R \in [0,\pi/2)$. As surfaces diverge, $\det(\mathbf{A})$ shrinks, spatial zeros spread further apart, coherence cells stretch, and fewer orthogonal modes physically fit the receiver surface.
\par \textit{Remark}: The mathematical identity $\det(\mathbf{A}^{-1}) = 1/\det(\mathbf{A})$ utilized in (\ref{Acohskewed}) is valid under the condition that the projection matrix $A$ is not singular, i.e., $\det(\mathbf{A}) \neq 0$. Based on (\ref{detA}) the determinant goes strictly to zero when either $\gamma_T = 90^\circ$ or $\gamma_R = 90^\circ$. Geometrically, this extreme edge-on configuration aligns a surface parallel to the propagation path, making the skewed coherence cell infinite and preventing spatial mode resolution. The proposed framework avoids this singularity: the visibility algorithm in Section IV evaluates signed vertex-to-plane distances and effective-centroid normal projections. As defined in Algorithm 1, the vertex-distance test (ALL $d_{Tx}(i,j) \le 0$ or ALL $d_{Rx}(i,j) \le 0$) and the effective-centroid guard $((\hat{\mathbf{n}}_T \cdot \mathbf{r}_{eff}) \le 0$ or $(\hat{\mathbf{n}}_R \cdot (-\mathbf{r}_{eff})) \le 0)$ rigorously classify these $90^\circ$ grazing limits as states of "No Visibility", safely bypassing the inversion of the undefined matrix.
\par
\begin{figure}
\centering
\includegraphics[width=0.45\textwidth]{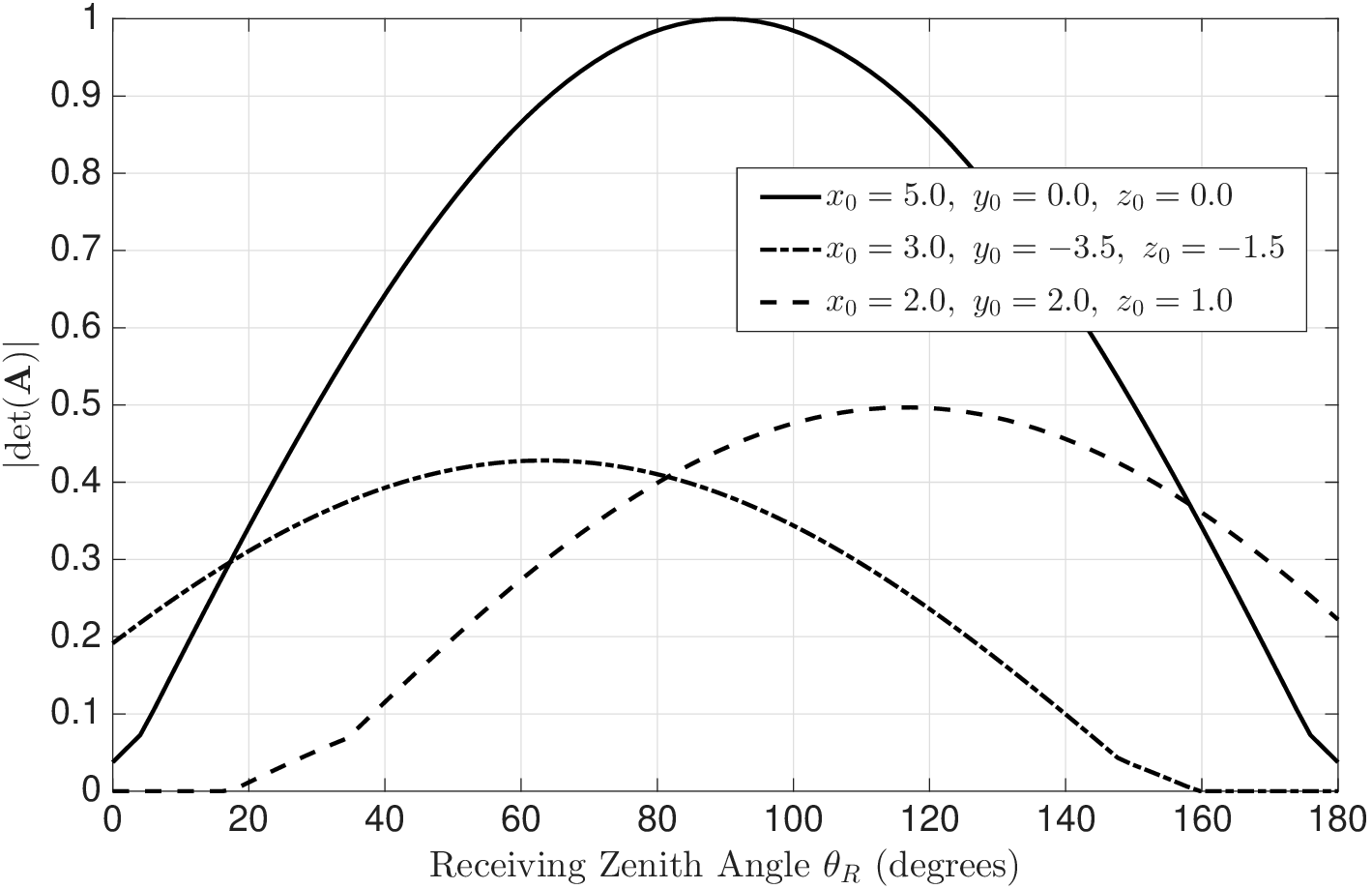}
\caption{The determinant of matrix $\mathbf{A}$ for various cases and for $L_{Tx}=0.75m$, $L_{Ty}=0.75m$, $L_{Rx}=0.25m$, $L_{Ry}=0.25m$, $\theta_T=90^o$, $\phi_T=0^o$, $\phi_R=180^o$.}
\label{Det_cases}
\end{figure}

 Fig.\ref{Det_cases} presents the value of the determinant for three indicative cases for the corresponding displacement of the Rx surface with respect to the Tx one. In all cases, the surfaces face each other, i.e., $\phi_T = 0^o$, $\phi_R = 180^o$, and the Tx surface is vertically placed. It is evident that in the paraxial placement the determinant takes its maximum value, whereas for any other displacement or tilt of the Rx surface smaller values are observed.
\begin{figure*}[!ht]
\centering
  \begin{subfigure}[t]{.31\linewidth}
  \includegraphics[trim=0 0 0 0,clip,width=\linewidth]{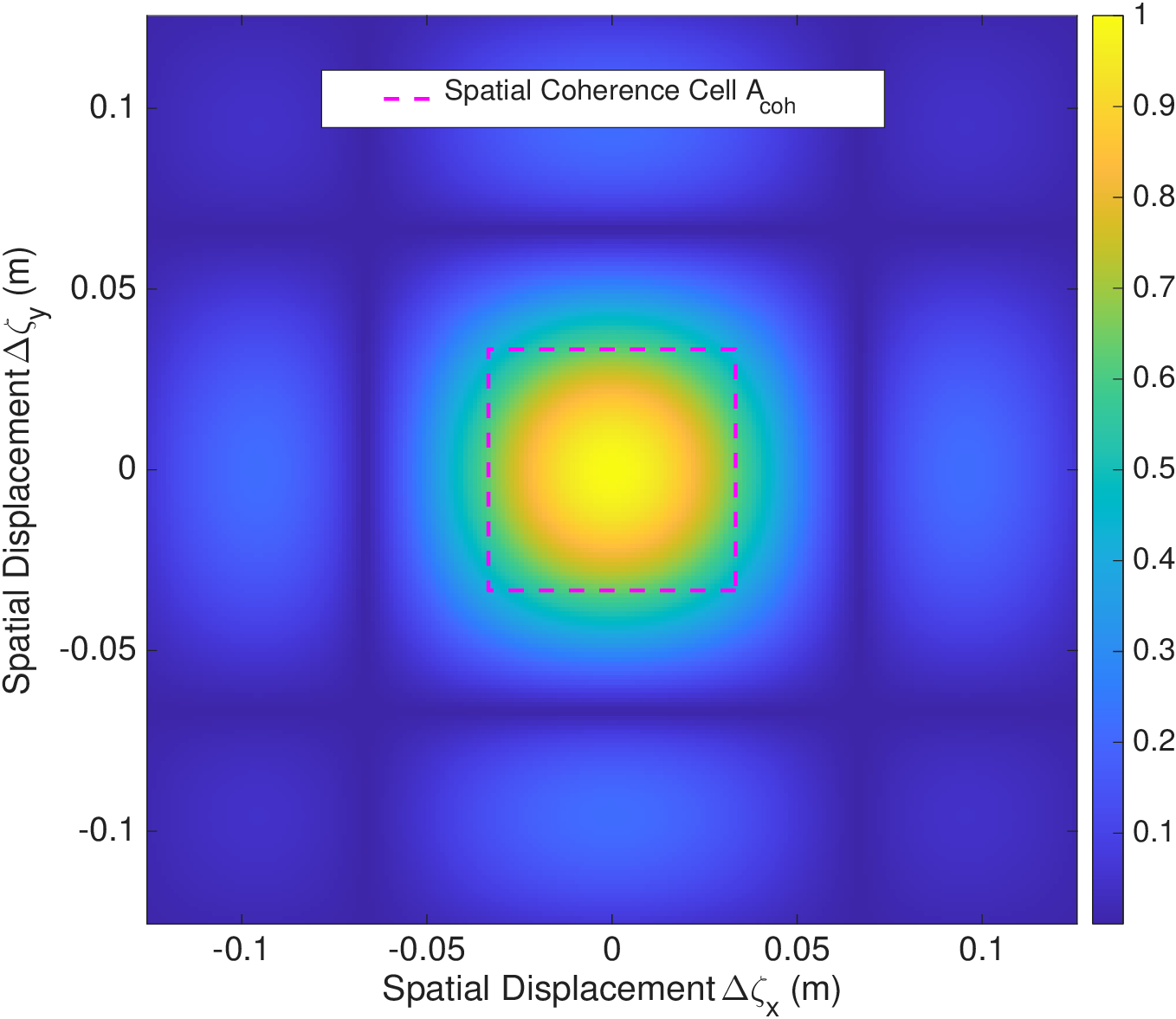}
\caption{Tx and Rx rotation angles, $\theta_T=90^o,\, \phi_T = 0^o, \, \theta_R=90^o, \, \phi_R=180^o$.}
  \label{Kern_Comp_1a}
  \end{subfigure}\hfil
  \begin{subfigure}[t]{.31\linewidth}
    \includegraphics[trim=0 0 0 0,clip,width=\linewidth]{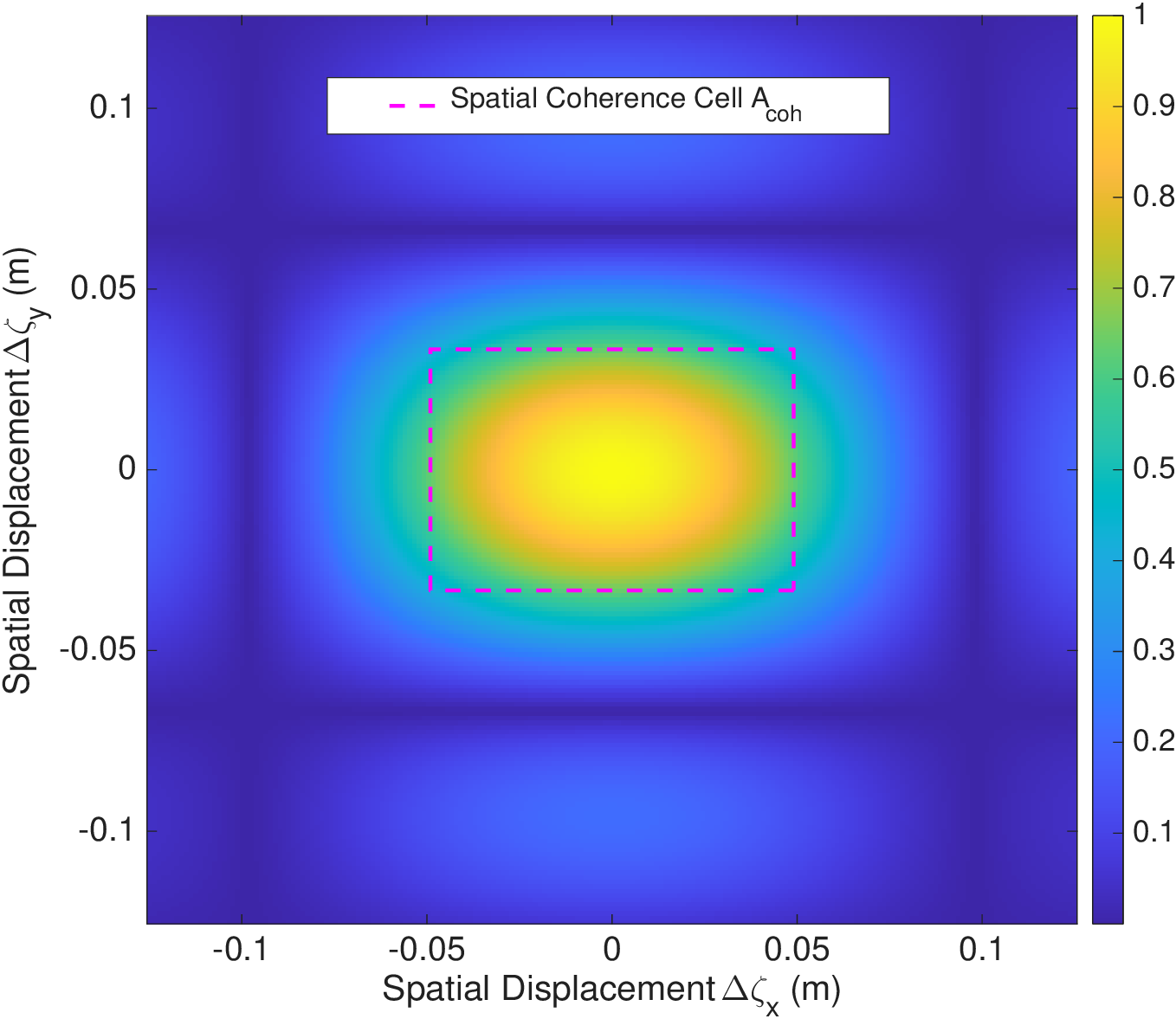}  \caption{Tx and Rx rotation angles, $\theta_T=90^o, \, \phi_T = 0^o, \, \theta_R=45^o, \, \phi_R=180^o$.}
    \label{Kern_Comp_1b}
  \end{subfigure}\hfil
    \begin{subfigure}[t]{.31\linewidth}
    \includegraphics[trim=0 0 0 0,clip,width=\linewidth]{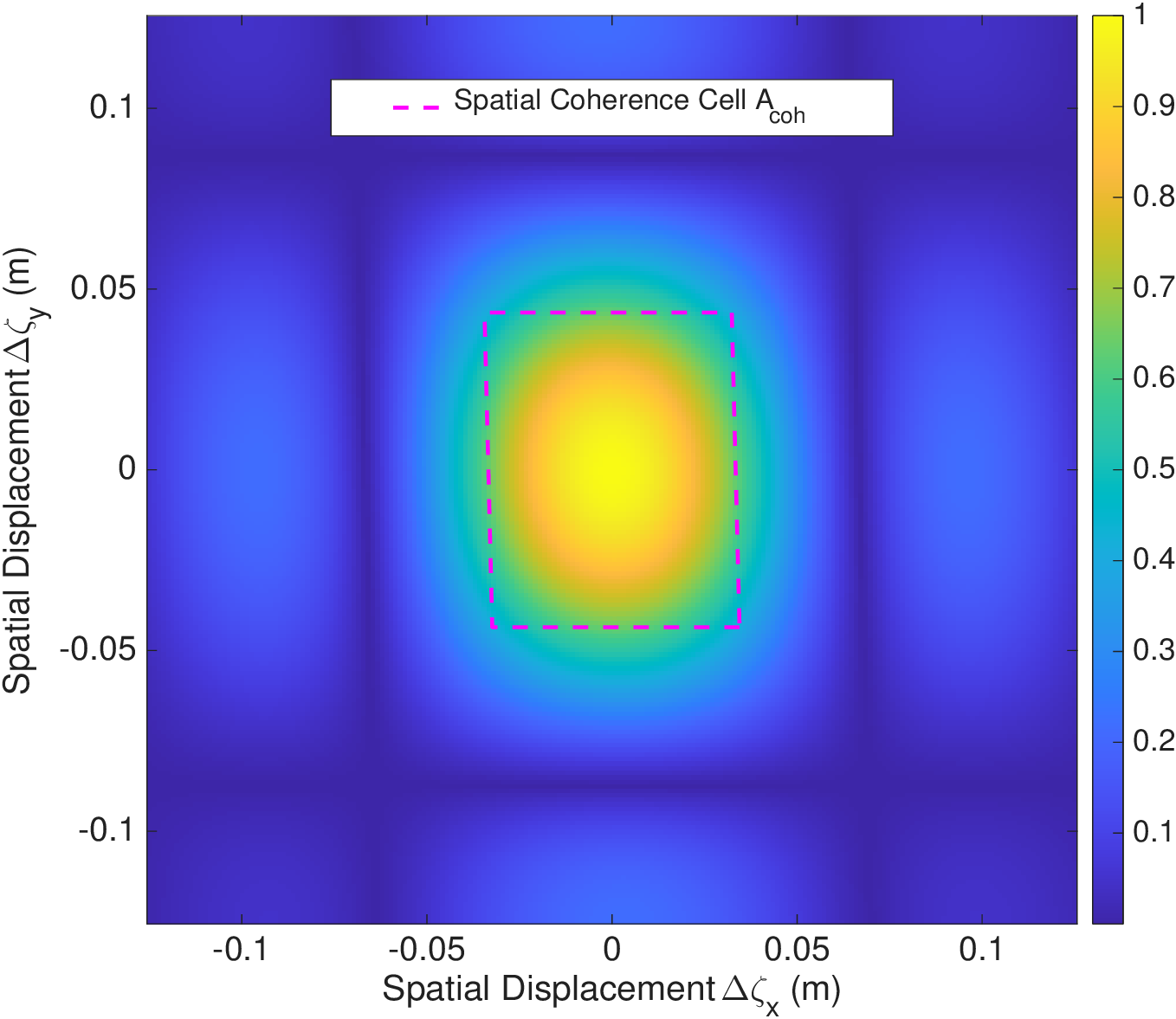}  \caption{Tx and Rx rotation angles, $\theta_T=90^o, \, \phi_T = 0^o, \, \theta_R=90^o, \, \phi_R=220^o$.}
    \label{Kern_Comp_1c}
  \end{subfigure}\hfil
  \\
\begin{subfigure}[t]{.31\linewidth}
  \includegraphics[trim=0 0 0 0,clip,width=\linewidth]{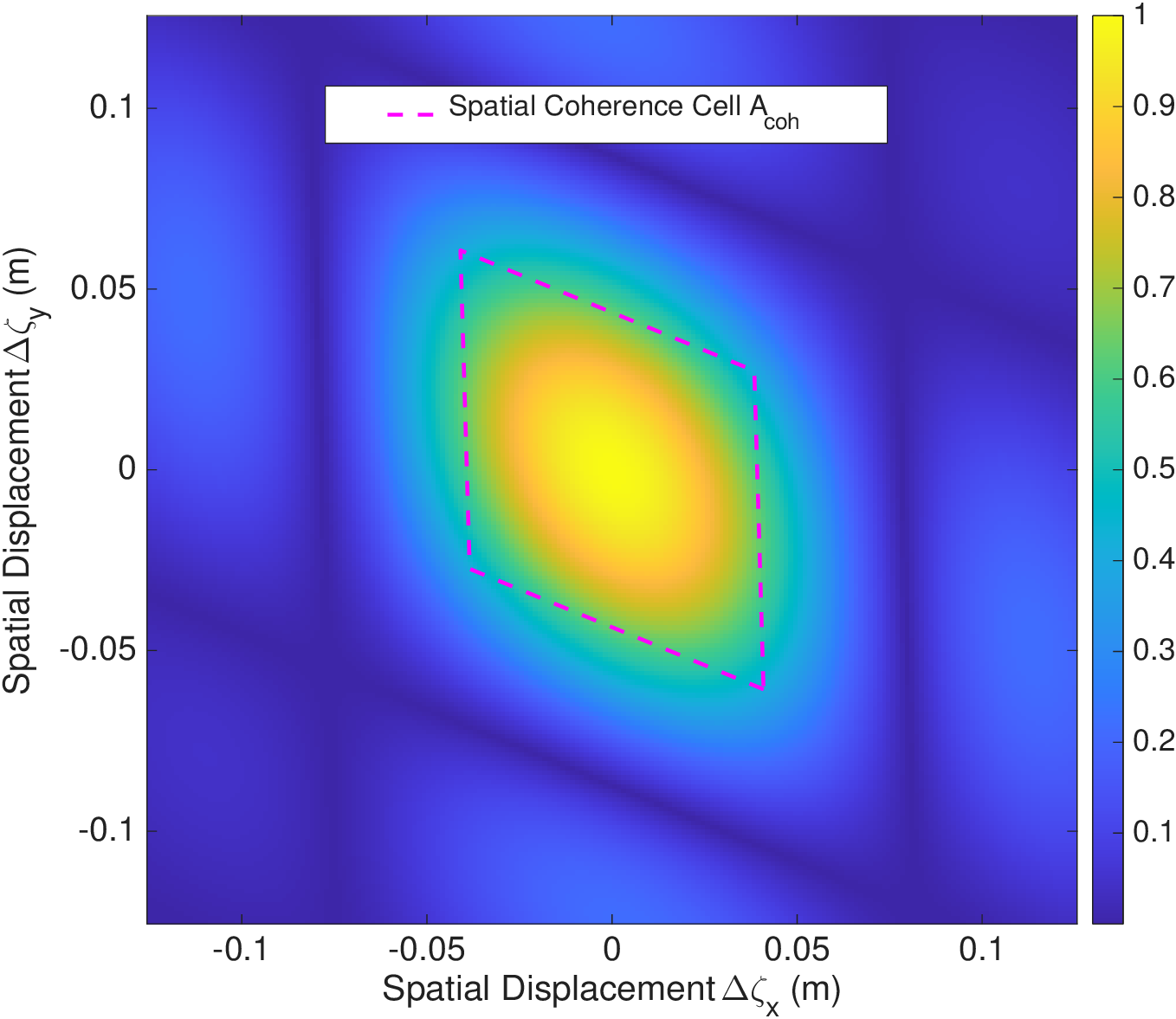}
   \caption{Tx and Rx rotation angles, $\theta_T=90^o, \, \phi_T = 0^o, \, \theta_R=60^o, \, \phi_R=220^o$.}
  \label{Kern_Comp_1d}
  \end{subfigure}\hfil
  \begin{subfigure}[t]{.31\linewidth}
    \includegraphics[trim=0 0 0 0,clip,width=\linewidth]{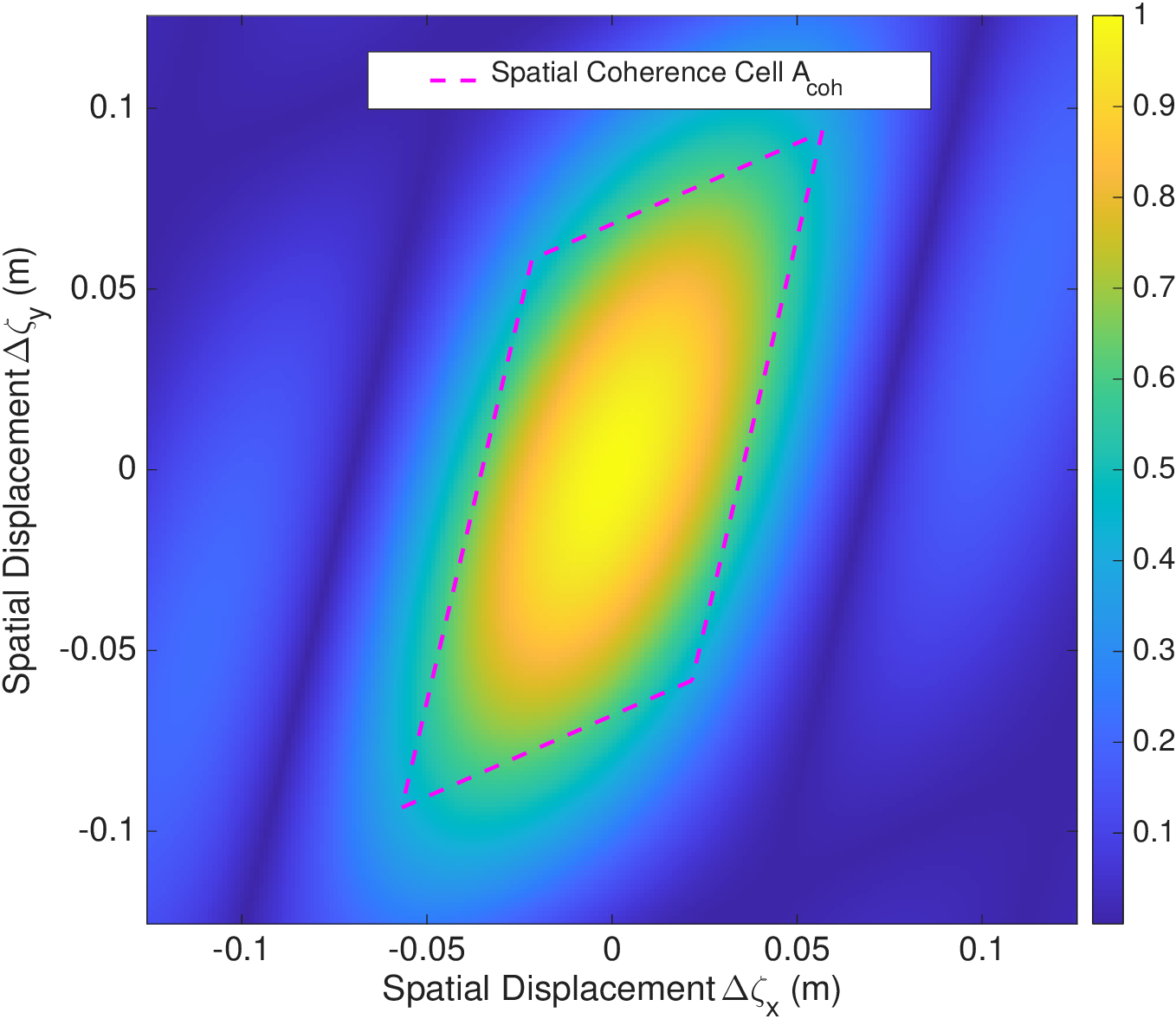}%
    \caption{Tx and Rx rotation angles, $\theta_T=90^o, \, \phi_T = 0^o, \, \theta_R=90^o, \, \phi_R=180^o$.}
    \label{Kern_Comp_1e}
  \end{subfigure}\hfil
    \begin{subfigure}[t]{.31\linewidth}
    \includegraphics[trim=0 0 0 0,clip,width=\linewidth]{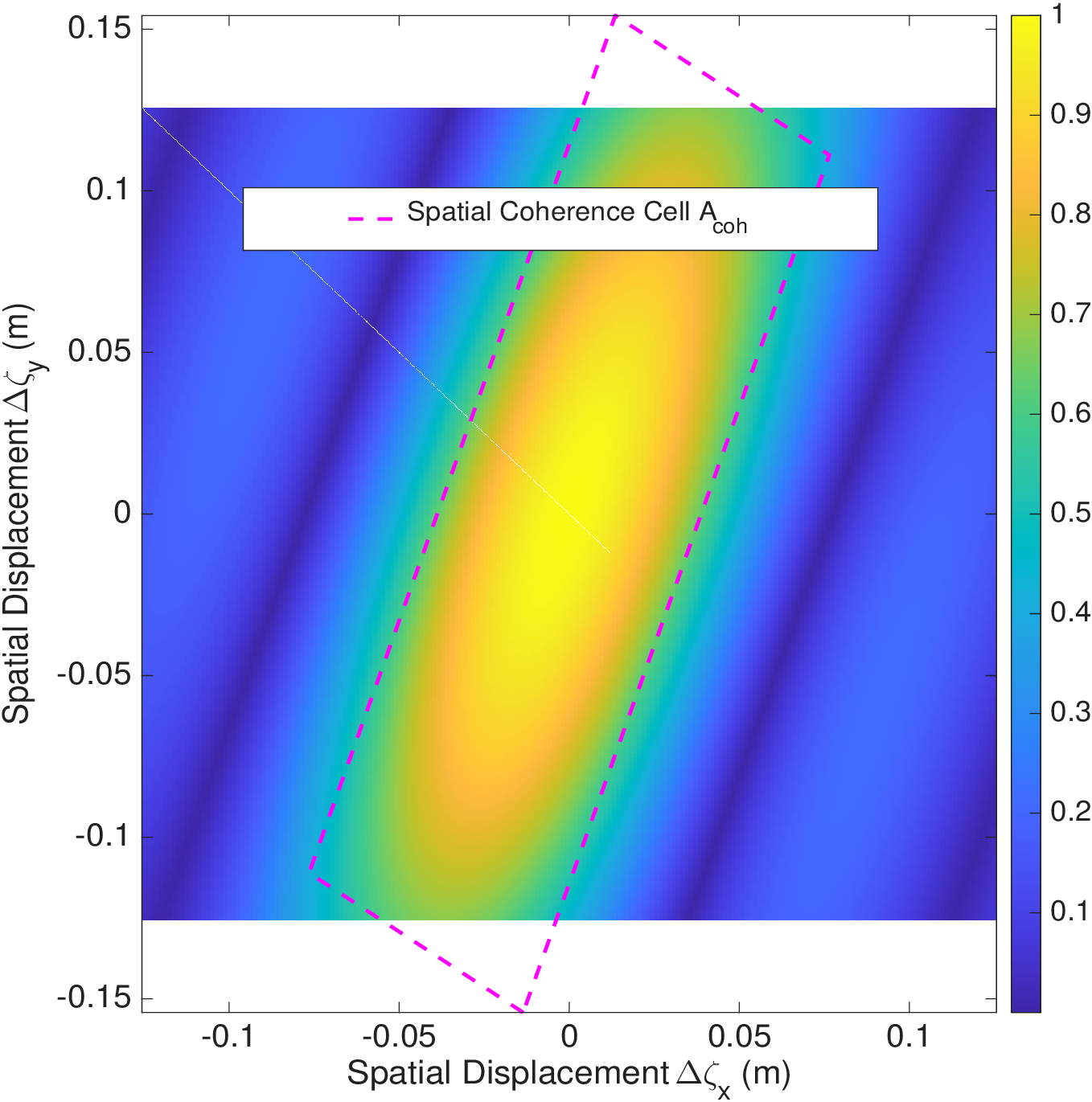}  \caption{Tx and Rx rotation angles, $\theta_T=90^o, \, \phi_T = 0^o, \, \theta_R=60^o, \, \phi_R=210^o$.}
    \label{Kern_Comp_1f}
  \end{subfigure}\hfil
  
     \caption{The magnitude of the kernel function on the Rx surface given by (\ref{Full_Kernel}) for six distinct cases. The size of the surfaces is $L_{Tx}=0.75\,m, L_{Ty}=0.75\,m, L_{Rx}=0.25\,m, L_{Ry}=0.25\,m,$. The center of the Rx is at $x_0=5\,m, \, y_0=0\,m, \, z_0=0\,m$, for the cases a), b) , c) and d), whereas for e) and f) at $x_0=3.0\,m, \, y_0=-3.5\,m, \, z_0=-1.5\,m$. The frequency is $f=30\,GHz$.}
  \label{Kernel_Comp_1}
\end{figure*}
Fig.\ref{Kernel_Comp_1} presents examples of the kernel function given by (\ref{Full_Kernel}) on the Rx surface, as well as the corresponding area of a spatial coherence cell. Case Fig.\ref{Kern_Comp_1a} depicts the kernel for perfectly aligned surfaces, whereas cases Fig.\ref{Kern_Comp_1b} and Fig.\ref{Kern_Comp_1c} depict the effect of rotation of the Rx in the zenith and azimuth angles, respectively. Fig.\ref{Kern_Comp_1d} shows how the double $\rm{sinc}$ function is skewed when the Rx is rotated in azimuth and zenith angles. In Fig.\ref{Kern_Comp_1e} the Rx is displaced, and in Fig.\ref{Kern_Comp_1f} a rotation in both zenith and azimuth is added, producing a highly skewed coherence cell extending out of the Rx surface. The fact that the cell has this extremely elongated shape is due to the projection matrix $\mathbf{A}$. As the surfaces tilt relative to each other, the determinant $\det(\mathbf{A})$ decreases, and the coherence cell becomes distorted and stretched. This partial fit of the cell on the Rx surface explains exactly why the calculation of the DoF produces fractional numbers and how they physically translate into the system. The DoF is calculated as the ratio of the Rx surface area to the area of the coherence cell, without requiring the cells to fit geometrically like perfect building blocks. In terms of the coupling intensity of orthogonal channels, the partial fit of a cell is translated to a lower coupling intensity and a consequent weaker spatial channel. These weak channels will suffer from higher attenuation and may require more power to be practically utilized as independent streams.
\par
\begin{figure*}[!ht]
\centering
  \begin{subfigure}[t]{.395\linewidth}
  \includegraphics[trim=0 0 0 0,clip,width=\linewidth]{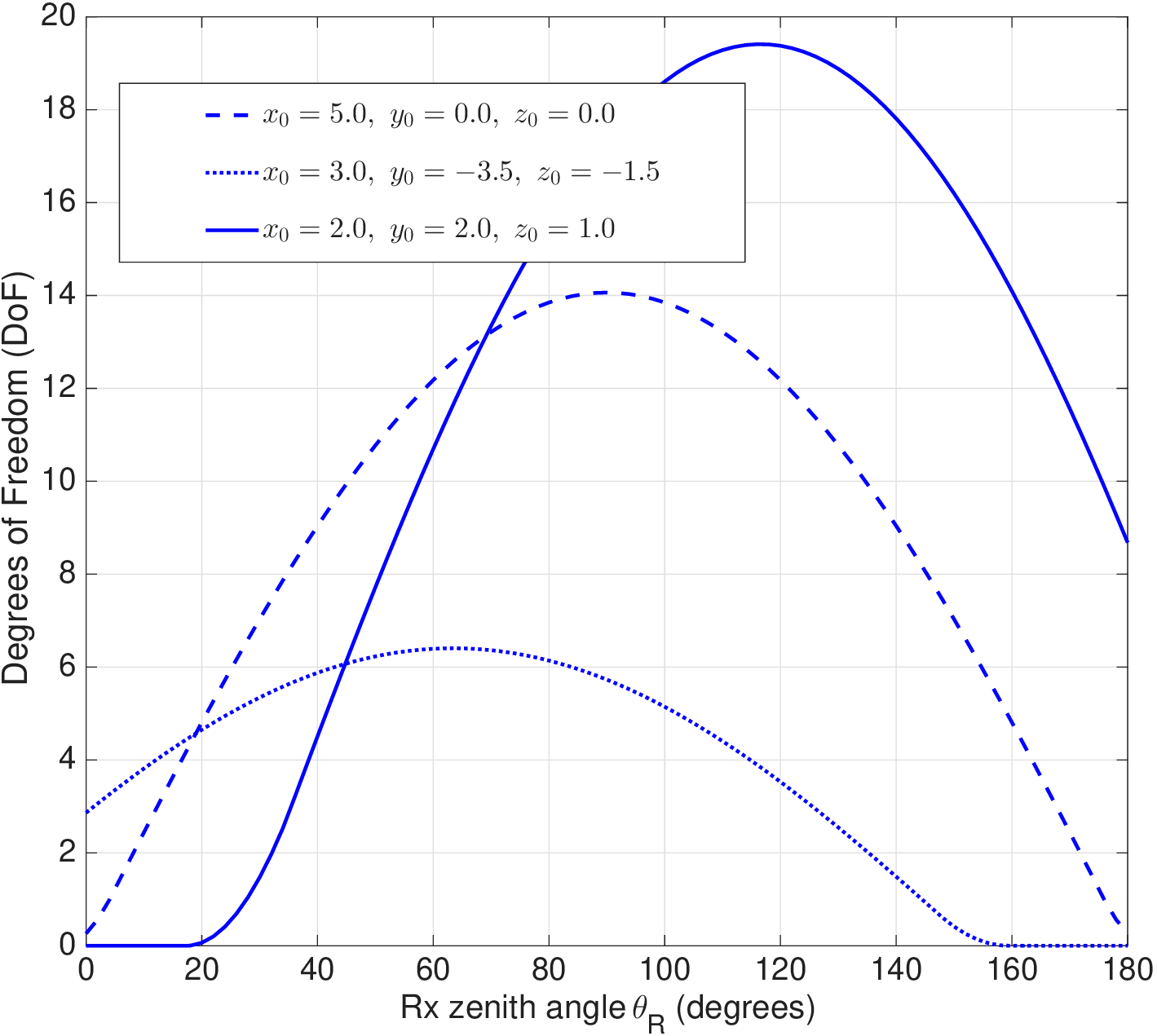}
\caption{Rotation angles, $\theta_T=90^o,\, \phi_T = 0^o, \, \phi_R=180^o$.}
  \label{DoF_Comp_1a}
  \end{subfigure}\hfil
  \begin{subfigure}[t]{.39\linewidth}
    \includegraphics[trim=0 0 0 0,clip,width=\linewidth]{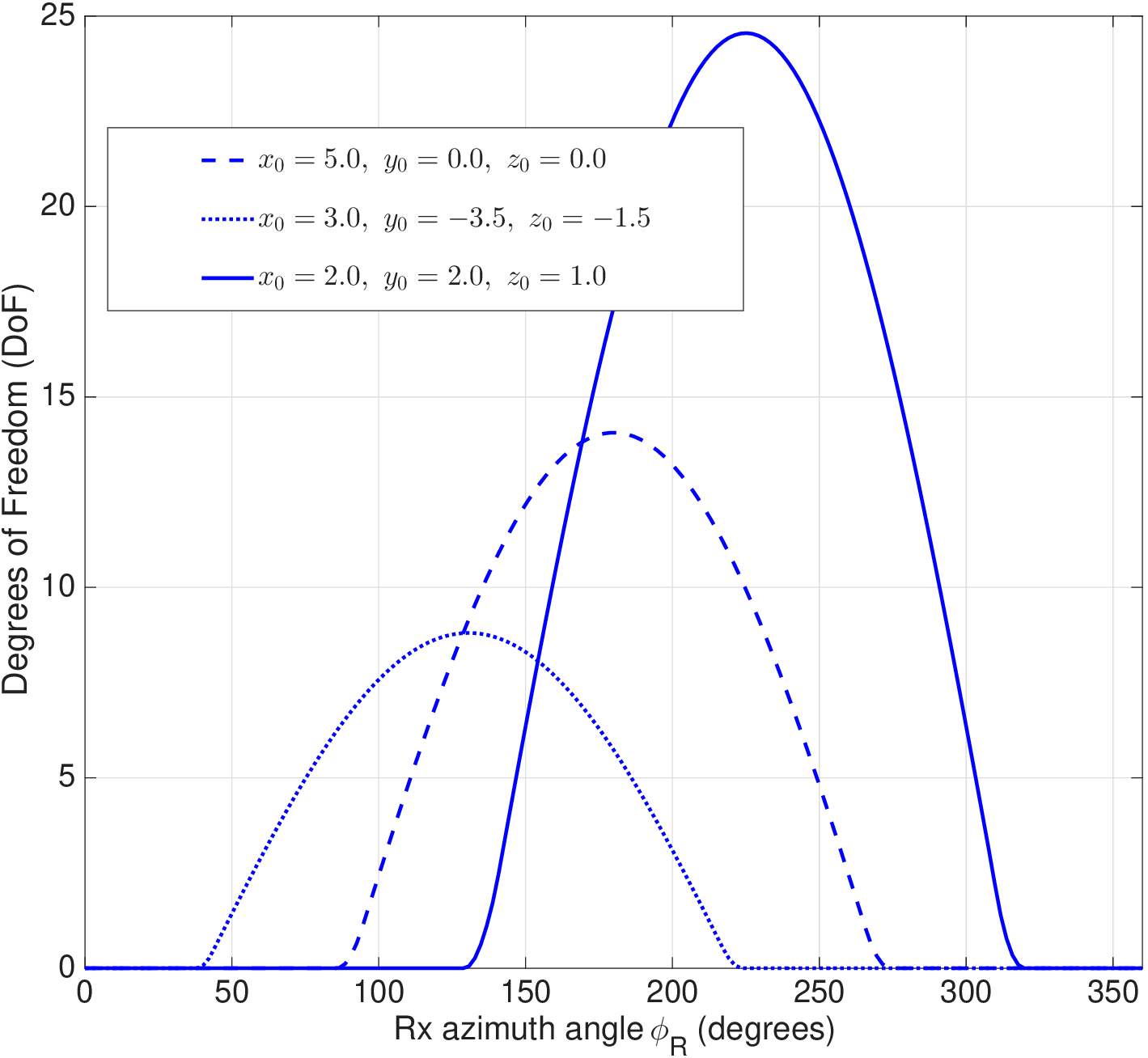}  \caption{Rotation angles, $\theta_T=90^o, \, \phi_T = 0^o, \, \theta_R=90^o.$}
    \label{DoF_Comp_1b}
  \end{subfigure}\hfil
     \caption{Degrees of Freedom (DoF) for three cases of the Rx center position. The size of the surfaces is $L_{Tx}=0.75\,m, L_{Ty}=0.75\,m, L_{Rx}=0.25\,m, L_{Ry}=0.25\,m,$. The frequency is $f=30\,GHz$.}
  \label{DoF_1}
\end{figure*}

 Fig.\ref{DoF_1} presents the DoF for three cases of the center position of the Rx surface. When the centers of the two surfaces are along the $x$-axis, i.e., a typical paraxial case, the DoF maximizes for $\theta_R = 90^o$ and $\phi_R = 180^o$ as expected. When Rx is displaced, the location of the maximum is also displaced. An interesting observation is related to the values of the determinant shown in Fig.\ref{Det_cases} and the distance of the two surface centers; while for the typical paraxial placement the determinant preserves greater values than those for the case $x_0=2.0, \, y_0=2.0, \, z_0=1.0, \, m$, the behavior in DoF is inverted due to the closer distance of the two centers, i.e. $d_0=5 \, m$ and $d_0=3 \, m$, respectively.
\subsubsection{DoF in Partial Visibility}
Under partial visibility, the critical difference is the integration domain in (\ref{kernel_2}), which is no longer a symmetric rectangle centered on the origin, but an $N$-sided clipped polygon $S_{T,eff}$, and now the integral over a generic 2D polygon does not separate into independent axes. To calculate the kernel, one may apply Green's theorem to transform the area integral over the polygon into a discrete sum of 1D contour integrals along its $N$ edges. Let edge $i$ be defined by its midpoint $\tilde{\mathbf{C}}_i$ and its full vector length $\Delta\mathbf{V}_i$. Then, if  $\hat{\mathbf{n}}_{P,i}$ is the unit outward normal vector for edge $i$, it is proved in Appendix C that the partial visibility kernel is given by (\ref{partialkernel}) shown at the bottom of next page.
\begin{figure*}[h!b]
	\hrulefill
\begin{equation}\label{partialkernel}
	\vert{}K_{partial}(\Delta\tilde{\zeta}_x, \Delta\tilde{\zeta}_y)\vert{} \propto \left\vert{} \frac{1}{\vert{}\tilde{\mathbf{K}}_\rho\vert{}^2} \sum_{i=1}^N (\hat{\mathbf{n}}_{P,i} \cdot \tilde{\mathbf{K}}_\rho) \vert{}\Delta\mathbf{V}_i\vert{} \exp(j \tilde{\mathbf{K}}_\rho \cdot \tilde{\mathbf{C}}_i) \text{sinc}\left( \frac{\tilde{\mathbf{K}}_\rho \cdot \Delta\mathbf{V}_i}{2\pi} \right) \right\vert{}
\end{equation}
\end{figure*}
The calculation of the kernel is possible through the algorithm provided in Section IV. However, the calculation of the DoF becomes quite demanding and for this reason an alternative method is proposed, based on the effective area of the spatial coherence cell on the receiver surface. This is defined by the integral of the normalized squared magnitude of the kernel as follows:
\begin{equation}\label{Acoh}
A_{coh} = \frac{\iint_{S_{R,eff}} \vert{}K_{partial}(\Delta\tilde{\zeta}_x, \Delta\tilde{\zeta}_y)\vert{}^2 \, {\rm{d}}(\Delta\tilde{\zeta}_x) \, {\rm{d}}(\Delta\tilde{\zeta}_y)}{\vert{}K_{partial}(0,0)\vert{}^2}.
\end{equation}
The resulting area acts as the fundamental quantum of spatial information. If the receiver has physical area $S_{R,eff}$, the exact number of independent data streams it can resolve is simply $S_{R,eff} / A_{coh}$. The kernel for zero spatial displacement in the denominator of (\ref{Acoh}) gives the peak value, that equals the area of the unblocked polygon $S_{T,eff}$. To evaluate the numerator, we recognize that the spatial kernel is the 2D Fourier transform of the transmitting aperture's indicator function, which is $1$ inside the polygon $P$ and $0$ outside. According to Parseval's theorem, the energy in the spatial domain must equal the energy in the frequency domain. Let the spatial frequency vector be $\mathbf{k} = \tilde{\mathbf{K}}_\rho = \frac{2\pi}{\lambda d_{eff}} \mathbf{A} \Delta\tilde{\boldsymbol{\zeta}}$. The change of variables from the spatial coordinates $\Delta\tilde{\boldsymbol{\zeta}}$ to the frequency coordinates $\mathbf{k}$ is:
\begin{equation}
{\rm{d}}^2\mathbf{k} = \left( \frac{2\pi}{\lambda d_{eff}} \right)^2 \left|\det(\mathbf{A})\right| \, {\rm{d}}^2(\Delta\tilde{\boldsymbol{\zeta}}).
\end{equation}
By Parseval's theorem:
$$\iint \vert{}K_{partial}\vert{}^2 \, {\rm{d}}^2\mathbf{k} = (2\pi)^2 \iint_{P} \vert{}1\vert{}^2 \, {\rm{d}}A = (2\pi)^2 S_{T,eff}$$
Substitute the Jacobian to switch the integral back to the physical receiver coordinates:
$$\left( \frac{2\pi}{\lambda d_{eff}} \right)^2 \left|\det(\mathbf{A})\right| \iint\limits_{S_{R,eff}} \vert{}K_{partial}\vert{}^2 \,{\rm{d}}^2(\Delta\tilde{\boldsymbol{\zeta}}) = (2\pi)^2 S_{T,eff}$$
where we have assumed that the diffraction mainlobe energy is well-contained within $S_{R,eff}$ and so the integral over $S_{R,eff}$ approximates the integral over $\mathbb{R}^2$. Solving for the integral:
$$\iint_{S_{R,eff}} \vert{}K_{partial}\vert{}^2 \, {\rm{d}}^2(\Delta\tilde{\boldsymbol{\zeta}}) = \frac{(\lambda d_{eff})^2}{\left|\det(\mathbf{A})\right|} S_{T,eff}$$
Substitute the numerator and the denominator back into (\ref{Acoh}):
\begin{equation}
A_{coh} = \frac{\frac{(\lambda d_{eff})^2}{|\det(\mathbf{A})|} S_{T,eff}}{(S_{T,eff})^2} = \frac{(\lambda d_{eff})^2}{|\det(\mathbf{A})| S_{T,eff}}.
\end{equation}
Finally, the total spatial DoF is given by:
\begin{equation}\label{DOFscalar_partial}
\text{DoF}_{scalar} = \frac{S_{R,eff}}{A_{coh}} = |\det(\mathbf{A})| \frac{S_{T,eff} S_{R,eff}}{(\lambda d_{eff})^2}.
\end{equation}
It is observed that the DoF is agnostic to the shape of the blockages on both surfaces. The spatial multiplexing capacity degrades strictly by the scalar loss of physical area at both ends, scaled by the 3D alignment penalty implied by $\det(\mathbf{A})$.
\par Interestingly, the same result is produced if the continuous channel operator $\mathcal{H}$ is used, defined as the linear integral operator mapping any square-integrable transmit source distribution to the received scalar field. In functional analysis, the number of significant DoF of a spectrally concentrated Hilbert-Schmidt operator is well-approximated by its effective rank, defined as the ratio of the squared Hilbert-Schmidt norm to its maximum eigenvalue \cite{Versh}:
\begin{equation}\label{newDOFscalar}
\text{DoF}_{scalar} \approx \frac{\Vert \mathcal{H} \Vert_{HS}^2}{\lambda_{max}}.
\end{equation}
The squared Hilbert-Schmidt norm is the double area integral of the squared magnitude of the path-loss-normalized scalar Green's function $\tilde{G}(\tilde{\eta}_x, \tilde{\eta}_y, \tilde{\zeta}_x, \tilde{\zeta}_y)$ over both partial surfaces:
\begin{equation}\label{HSnorm}
	\begin{aligned}
\Vert \mathcal{H} \Vert_{HS}^2 & = \iint\limits_{S_{R,eff}} \iint\limits_{\,\,\,S_{T,eff}} \vert{}\tilde{G}(\tilde{\pmb{\eta}}, \tilde{\pmb{\zeta}})\vert{}^2 \, {\rm{d}}\tilde{\pmb{\eta}} \, {\rm{d}}\tilde{\pmb{\zeta}}  = S_{R,eff} S_{T,eff}
	\end{aligned}
\end{equation}
since we have ignored the path loss constants and kept only the exponential term of the channel. The maximum eigenvalue $\lambda_{max}$ is calculated by applying the Szegö's theorem \cite{Gray} to the power spectral density, which in turn is calculated as the Fourier transform of the kernel function in (\ref{kernel_2}), using the Jacobian determinant between the physical spatial coordinates and the spatial frequency coordinates. Therefore,
\begin{equation}\label{lmax}
	\lambda_{max} = \frac{(\lambda d_{eff})^2}{|{\rm{det}}(\mathbf{A})|}.
\end{equation}
Substituting (\ref{HSnorm}) and (\ref{lmax}) in (\ref{newDOFscalar}), then (\ref{DOFscalar_partial}) results.
\par
\begin{figure*}[!ht]
\centering
  \begin{subfigure}[t]{.33\linewidth}
  \includegraphics[trim=0 0 0 0,clip,width=\linewidth]{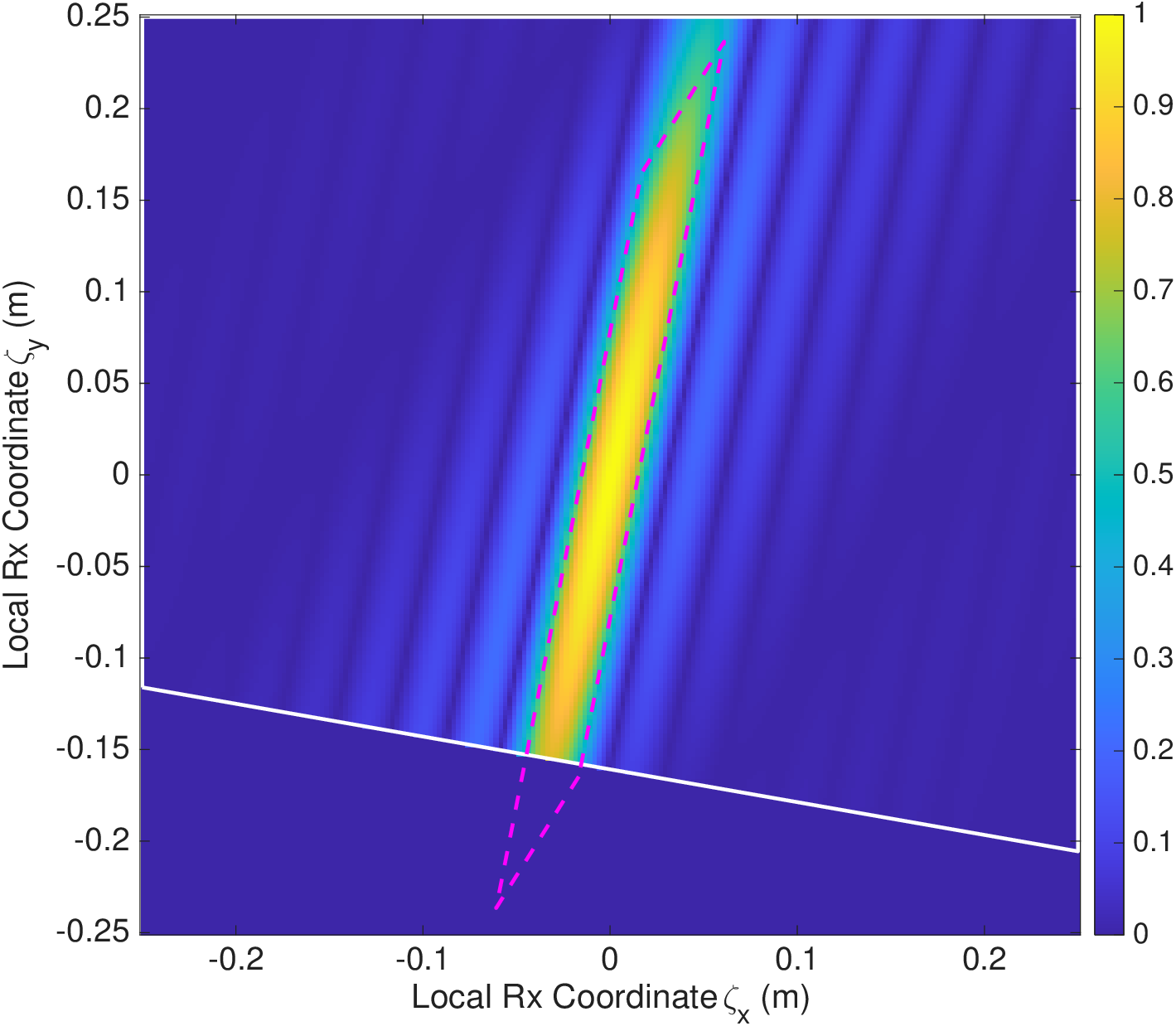}
\caption{Amplitude of kernel function for $\theta_T=100^o,\, \phi_T = -40^o, \, \theta_R=90^o, \, \phi_R=220^o$.}
  \label{partial_vis_a}
  \end{subfigure}\hfil
  \begin{subfigure}[t]{.33\linewidth}
    \includegraphics[trim=0 0 0 0,clip,width=\linewidth]{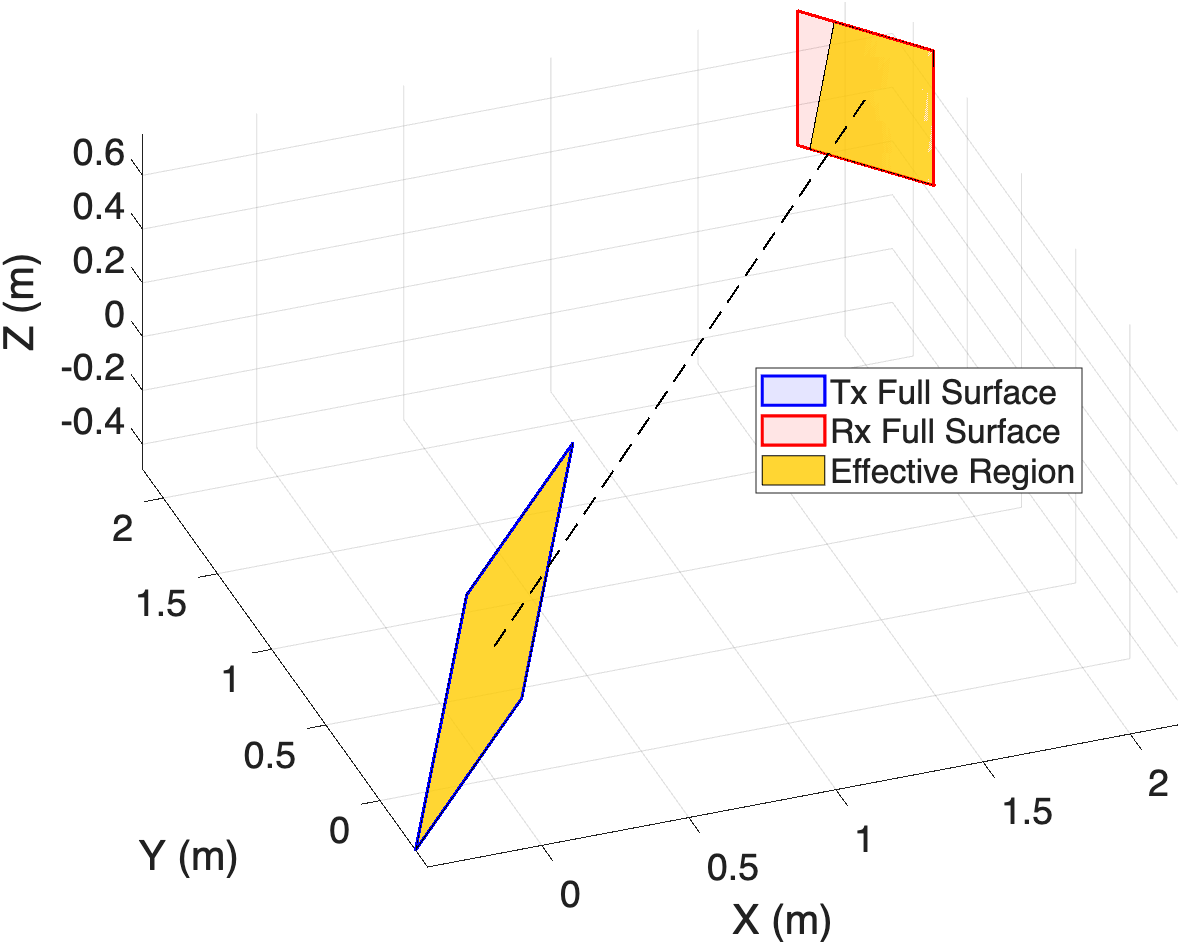}  \caption{Relative geometry for $\theta_T=100^o,\, \phi_T = -40^o, \, \theta_R=90^o, \, \phi_R=220^o$.}
    \label{partial_vis_b}
  \end{subfigure}\hfil
  \begin{subfigure}[t]{.31\linewidth}
    \includegraphics[trim=0 0 0 0,clip,width=\linewidth]{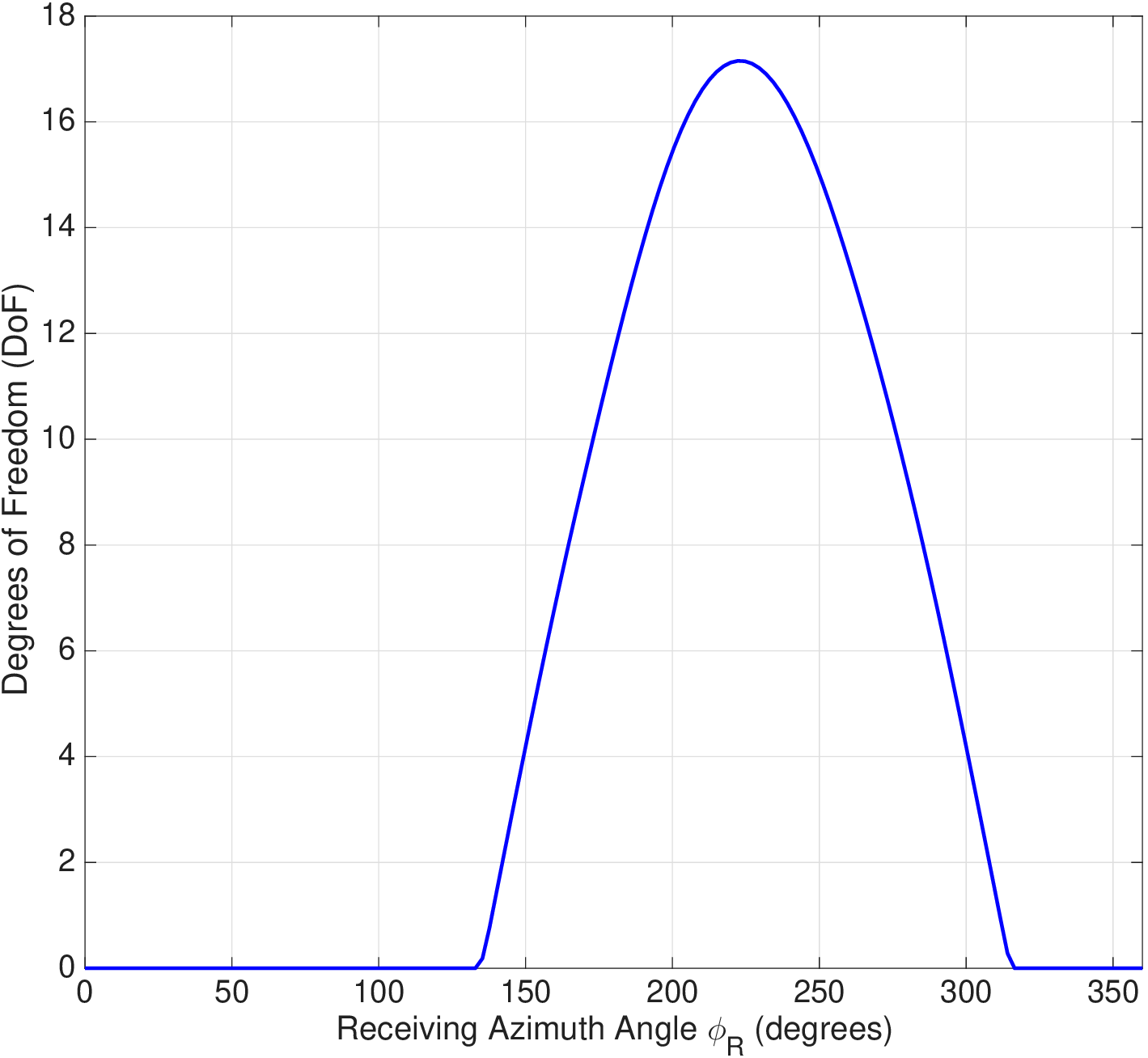}  \caption{DoF vs. $\phi_R$ for $\theta_T=100^o, \, \phi_T = -40^o, \, \theta_R=90^o.$}
    \label{partial_vis_c}
  \end{subfigure}\hfil
  \caption{A geometry case with two surfaces $L_{Tx}=1.0\,m, \, L_{Ty}=1.0\,m, \, L_{Rx}=0.5\,m, \, L_{Ry}=0.5\,m,$ at $f=30\,GHz$, and $\theta_T=100^o,\, \phi_T = -40^o$, that includes partial visibility. a) Amplitude of the kernel function and the coherence cell for the geometry shown in b). c) DoF vs. $\phi_R$.}
  \label{partial_vis}
\end{figure*}

 Fig.\ref{partial_vis_a} presents the amplitude of the kernel function and the spatial coherence cell on the Rx surface for a partial visibility condition shown in Fig.\ref{partial_vis_b}. A portion of the Rx surface is not visible and this is depicted with a dark blue color in the figure. The visible region is depicted by a white polygon. Moreover, the DoF vs. the rotation azimuth angle $\phi_R$ is calculated in Fig.\ref{partial_vis_c} using (\ref{DOFscalar_partial}) and the algorithm given in section IV.
\subsection{Vector Electromagnetic Fields}
\begin{figure*}[!ht]
\centering
  \begin{subfigure}[t]{.33\linewidth}
  \includegraphics[trim=0 0 0 0,clip,width=\linewidth]{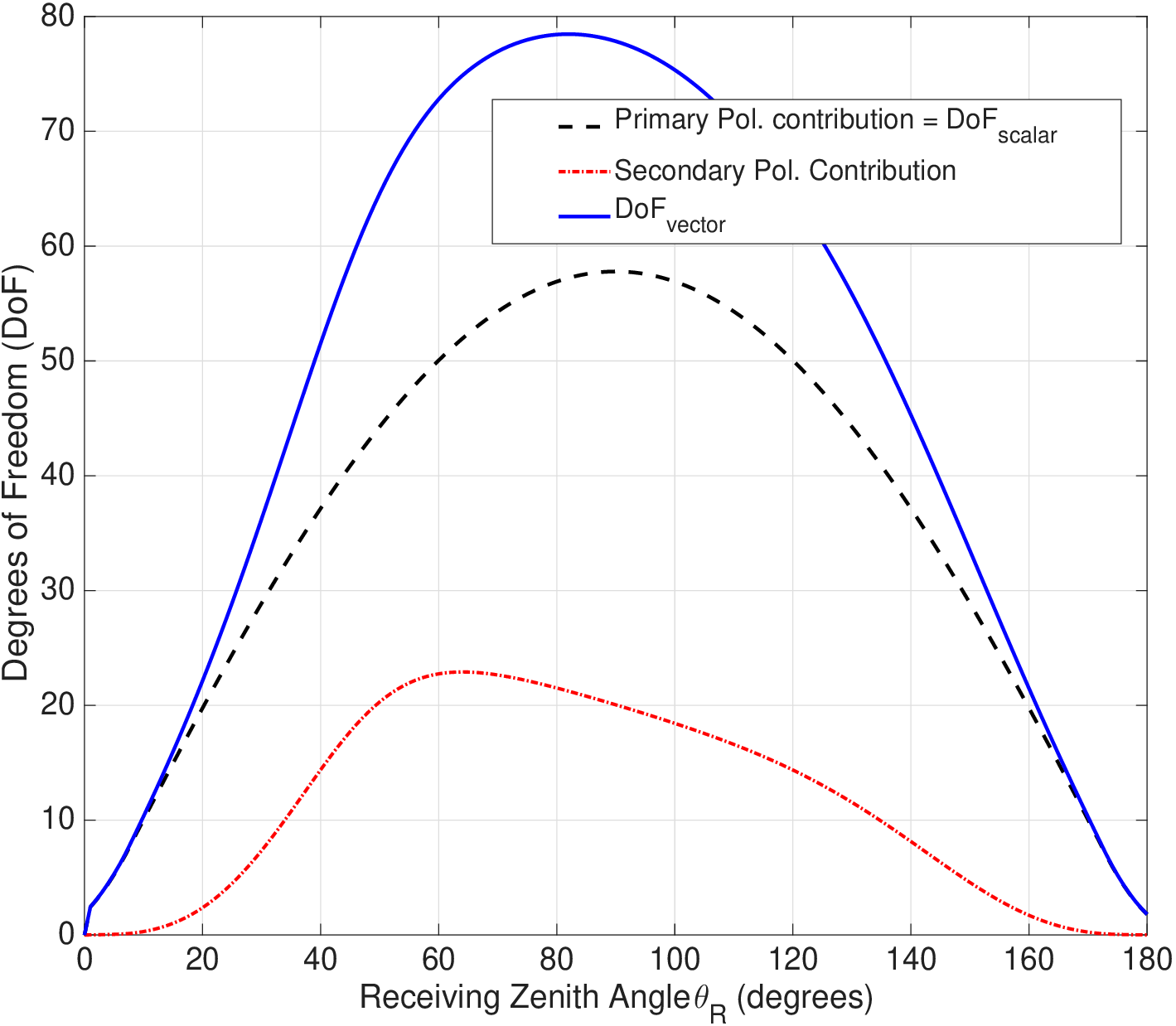}
\caption{$\theta_T=100^o,\, \phi_T = -40^o, \, \phi_R=220^o$.}
  \label{DoF_vec_1}
  \end{subfigure}\hfil
  \begin{subfigure}[t]{.33\linewidth}
    \includegraphics[trim=0 0 0 0,clip,width=\linewidth]{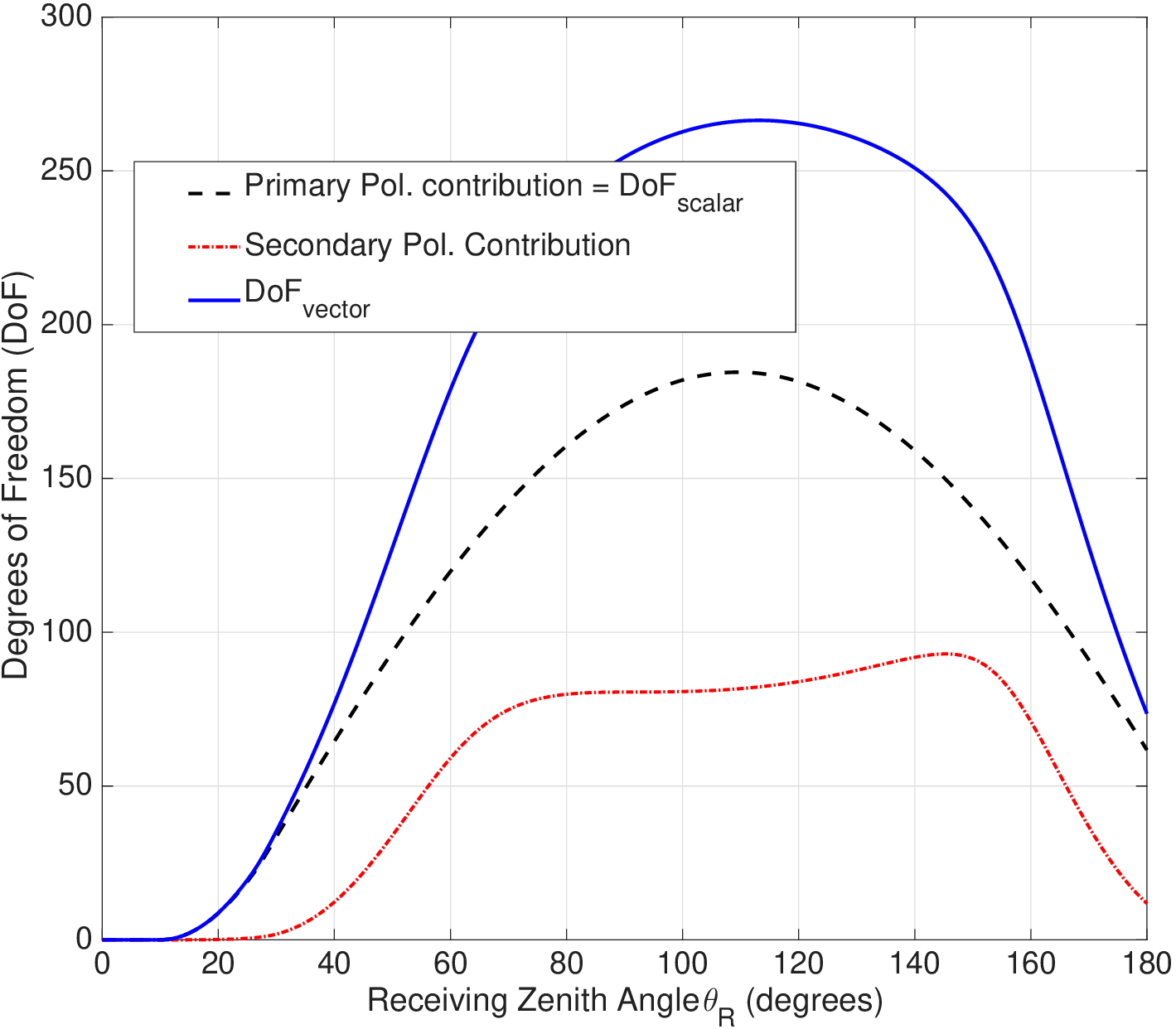}  \caption{$\theta_T=90^o,\, \phi_T = 0^o, \, \phi_R=220^o$.}
    \label{DoF_vec_2}
  \end{subfigure}\hfil
  \begin{subfigure}[t]{.32\linewidth}
    \includegraphics[trim=0 0 0 0,clip,width=\linewidth]{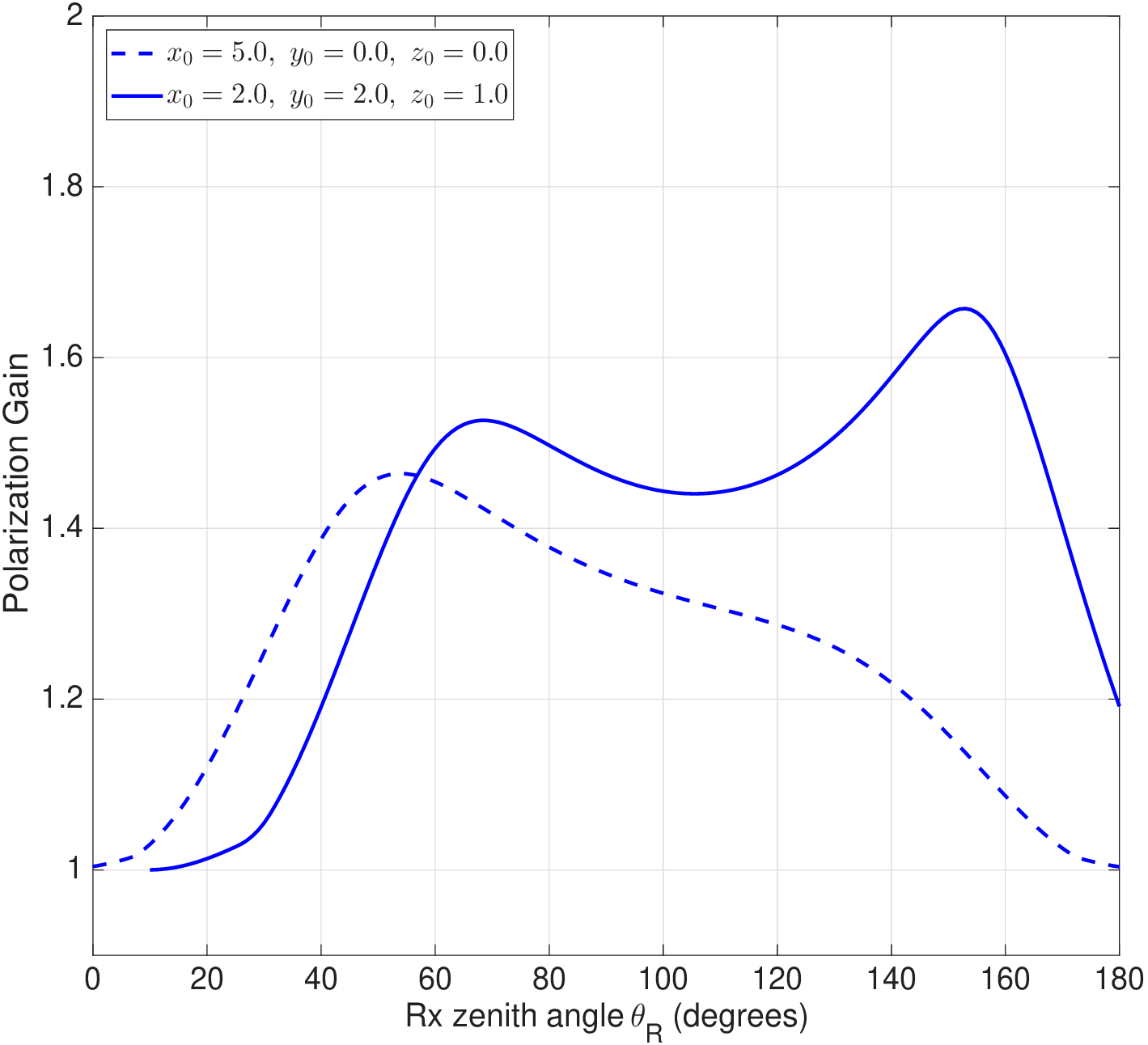}  \caption{Polarization gain}
    \label{Pol_gain}
  \end{subfigure}\hfil
  \caption{The vector fields DoF vs. the zenith angle $\theta_R$ for two surfaces $L_{Tx}=1.0\,m, \, L_{Ty}=1.0\,m, \, L_{Rx}=0.5\,m, \, L_{Ry}=0.5\,m,$ at $f=30\,GHz$. In a) the Rx center is located at $x_0=5,\,y_0=0,\,z_0=0$, and in b) at $x_0=2,\,y_0=2,\,z_0=1$. c) presents the polarization gain $\mathrm{tr}(X)/\mu_{\max} = 1 + \mu_2/\mu_1$, for the two cases.}
  \label{DoF_vec_fig}
\end{figure*}

Transitioning from scalar acoustics to full electromagnetic vector fields requires accounting for transverse waves and explicit polarization current alignment on the surfaces. The scalar Green's function $G(\pmb{\zeta}, \pmb{\eta})$ is replaced by the Dyadic Green's function $\overline{\overline{\mathbf{G}}}(\pmb{\zeta}, \pmb{\eta})$. In free space, the electric field generated by a vector current density is governed by:
\begin{equation}
\overline{\overline{\mathbf{G}}}(\pmb{\zeta}, \pmb{\eta}) = \left( \mathbf{I} + \frac{1}{k^2} \nabla \nabla \right) G(\pmb{\zeta}, \pmb{\eta}).
\end{equation}
In the radiative near-field and far-field, where $k\Vert{}\pmb{\zeta} - \pmb{\eta}\Vert{} \gg 1$, the gradient operator simplifies to a projection along the direction of propagation. Substituting the unit reference vector $\hat{\mathbf{r}}_{eff}$, the dyadic operator degenerates into a purely geometric transverse projection matrix:
\begin{equation}
\overline{\overline{\mathbf{G}}} (\pmb{\zeta}, \pmb{\eta}) \approx \left( \mathbf{I} - \hat{\mathbf{r}}_{eff}\hat{\mathbf{r}}_{eff}^T \right) G(\pmb{\zeta}, \pmb{\eta}).
\end{equation}
Physically, the projection matrix $\left( \mathbf{I} - \hat{\mathbf{r}}_{eff}\hat{\mathbf{r}}_{eff}^T \right)$ enforces the rule that electromagnetic waves cannot support longitudinal field components.
A continuous dual-polarized Tx surface supports independent current densities along its two local unit vectors ($\hat{\mathbf{u}}_{Tx}, \hat{\mathbf{u}}_{Ty}$). The Rx surface captures these fields by projecting them onto its own local unit vectors ($\hat{\mathbf{u}}_{Rx}, \hat{\mathbf{u}}_{Ry}$).
The vector channel gain between a specific transmit polarization $j \in \{x,y\}$ and a receive polarization $i \in \{x,y\}$ is dictated by the polarization coupling coefficient $\chi_{ij}$:
\begin{equation}\label{polcoef}
	\begin{aligned}
\chi_{ij} & = \hat{\mathbf{u}}_{Ri}^T \left( \mathbf{I} - \hat{\mathbf{r}}_{eff}\hat{\mathbf{r}}_{eff}^T \right) \hat{\mathbf{u}}_{Tj} \\ & = \hat{\mathbf{u}}_{Ri} \cdot \hat{\mathbf{u}}_{Tj} - (\hat{\mathbf{r}}_{eff} \cdot \hat{\mathbf{u}}_{Ri})(\hat{\mathbf{r}}_{eff} \cdot \hat{\mathbf{u}}_{Tj})
	\end{aligned}	.
\end{equation}
This geometric scalar scales the amplitude of the field. The resulting projected dyadic channel reduces the $3\times3$ dyadic Green's function in 3D space down to a $2 \times 2$ matrix confined to the local 2D planes of the transmitting and receiving surfaces. Actually, two projections take place; the first projects out any longitudinal field components along the direction of propagation, and the second is a surface-plane projection. Using the two local coordinate bases, $\mathbf{U}_T,\mathbf{U}_R$, defined in Appendix B, we can write for the projected dyadic channel
\begin{equation}
 \overline{\overline{\mathbf{G}}}_{\text{proj}} (\pmb{\zeta}, \pmb{\eta}) = \underbrace{\mathbf{U}_R^T\left( \mathbf{I} - \hat{\mathbf{r}}_{eff}\hat{\mathbf{r}}_{eff}^T \right)\mathbf{U}_T}_\mathcal{X} G(\pmb{\zeta}, \pmb{\eta})   
\end{equation}
where the elements of the matrix $\mathcal{X}$ are the polarization coupling coefficients $\chi_{ij}$ given in (\ref{polcoef}). Notice that the phase propagation, governed by the scalar $G(\pmb{\zeta}, \pmb{\eta})$ and its Taylor expansion, remains mathematically identical to the scalar analysis.
Because the surfaces are dual-polarized, the scalar kernel $K(\tilde{\pmb{\zeta}}, \tilde{\pmb{\zeta}}')$ expands into a $2 \times 2$ block tensor kernel $\overline{\overline{\mathbf{K}}}(\tilde{\pmb{\zeta}}, \tilde{\pmb{\zeta}}')$. This tensor captures not only the spatial covariance between different observation points, but also the cross-polarization interference (XPI) between orthogonal currents:
\begin{equation}
\begin{aligned}
\overline{\overline{\mathbf{K}}}(\tilde{\pmb{\zeta}}, \tilde{\pmb{\zeta}}') & = \begin{bmatrix} K_{xx}(\tilde{\pmb{\zeta}}, \tilde{\pmb{\zeta}}') & K_{xy}(\tilde{\pmb{\zeta}}, \tilde{\pmb{\zeta}}') \\ K_{yx}(\tilde{\pmb{\zeta}}, \tilde{\pmb{\zeta}}') & K_{yy}(\tilde{\pmb{\zeta}}, \tilde{\pmb{\zeta}}') \end{bmatrix} \\ &=\iint_{S_{T,\text{eff}}} \overline{\overline{\mathbf{G}}}_{\text{proj}}(\tilde{\pmb{\zeta}}, \tilde{\pmb{\eta}}) \, \overline{\overline{\mathbf{G}}}_{\text{proj}}^H(\tilde{\pmb{\zeta}}', \tilde{\pmb{\eta}}) \, d\tilde{\eta}_x d\tilde{\eta}_y \\ & = (\mathcal{X} \mathcal{X}^H) \iint_{S_{T,\text{eff}}} G(\tilde{\pmb{\zeta}}, \tilde{\pmb{\eta}}) G^*(\tilde{\pmb{\zeta}}', \tilde{\pmb{\eta}}) \, d\tilde{\eta}_x d\tilde{\eta}_y.
\end{aligned}
\end{equation}\label{Gram}
Next we define the Gram matrix $\mathbf{X} = \mathcal{X} \mathcal{X}^H$ as the polarization covariance matrix, with elements
\begin{equation}
    \mathbf{X} = \begin{bmatrix} \vert{}\chi_{xx}\vert{}^2 + \vert{}\chi_{xy}\vert{}^2 & \chi_{xx}\chi_{yx}^* + \chi_{xy}\chi_{yy}^* \\ \chi_{yx}\chi_{xx}^* + \chi_{yy}\chi_{xy}^* & \vert{}\chi_{yx}\vert{}^2 + \vert{}\chi_{yy}\vert{}^2 \end{bmatrix}.
\end{equation}
Assuming the polarization coupling coefficients $\chi_{ij}$ vary slowly across the array, the tensor kernel is simply given by the Kronecker product of the Gram matrix $\mathbf{X}$ and the original scalar spatial kernel $K_{scalar}$:
\begin{equation}
\overline{\overline{\mathbf{K}}}(\tilde{\pmb{\zeta}}, \tilde{\pmb{\zeta}}') \approx \mathbf{X} \otimes K_{scalar}(\tilde{\pmb{\zeta}}, \tilde{\pmb{\zeta}}').
\end{equation}
Following similar steps for the continuous vector channel operator $\overline{\overline{\mathcal{H}}}$ we may write 
\begin{equation}
\begin{aligned}
\Vert{}\overline{\overline{\mathcal{H}}}\Vert{}_{HS}^2 &= \iint_{S_{R,\text{eff}}} \iint_{S_{T,\text{eff}}} \left\Vert{}\overline{\overline{\mathbf{G}}}_{\text{proj}}(\tilde{\pmb{\zeta}}, \tilde{\pmb{\eta}})\right\Vert{}_F^2 \, d\tilde{\eta}_x d\tilde{\eta}_y \, d\tilde{\zeta}_x d\tilde{\zeta}_y \\
    & = {\rm{tr}}(\mathbf{X})\Vert{}\mathcal{H}\Vert{}_{HS}^2 =  {\rm{tr}}(\mathbf{X})S_{T,eff}S_{R,eff}.
\end{aligned}
\end{equation}
The DoF is then given by
\begin{equation}\label{DOFvector}
	\text{DoF}_{vector} = \frac{{\rm{tr}} (\mathbf{X})}{\mu_{max}} |\det(\mathbf{A})| \frac{S_{T,eff} S_{R,eff}}{(\lambda d_{eff})^2}.
\end{equation}\label{DoF_vec}
where $\mu_{max}$ is the maximum eigenvalue of $\mathbf{X}$. The eigenvalues $\mu_1=\mu_{max}, \mu_2$ of matrix $\mathbf{X}$ represent the power coupling gains of the two orthogonal polarization eigenmodes. It is easily shown that $\frac{{\rm{tr}} (\mathbf{X})}{\mu_{max}}=\frac{\mu_1 +\mu_2}{\mu_1}=1+\frac{\mu_2}{\mu_1}$, and thus,
\begin{equation}
  \text{DoF}_{vector} = \left(1+\frac{\mu_2}{\mu_1}\right) |\det(\mathbf{A})| \frac{S_{T,eff} S_{R,eff}}{(\lambda d_{eff})^2}  .
\end{equation}
Note that the normalized eigenvalues ($1$ and $\mu_2/\mu_1$) are the DoF weights per polarization. For opposing parallel surfaces, $\mu_1=\mu_2=1$, so the two polarizations double the DoF. Fig.\ref{DoF_vec_fig} presents two cases of the DoF when two polarizations are assumed. Fig.\ref{Pol_gain} shows the corresponding polarization gain.
\section{Visibility Algorithm}
This section presents an algorithm for the calculation of the DoF, based on the centroids of the clipped surfaces and the corresponding effective regions for the partial visibility case between two planar surfaces radiating in a single half-space. The algorithm utilizes the definition of the four corner vertices of the Tx and Rx surfaces as follows: 
\begin{equation}
	\begin{aligned}
& \mathbf{T}_{i,j} = \pm \frac{L_{Tx}}{2}\hat{\mathbf{u}}_{Tx} \pm \frac{L_{Ty}}{2}\hat{\mathbf{u}}_{Ty} \\
& \mathbf{R}_{i,j} = \mathbf{R}^0 \pm \frac{L_{Rx}}{2}\hat{\mathbf{u}}_{Rx} \pm \frac{L_{Ry}}{2}\hat{\mathbf{u}}_{Ry}	
	\end{aligned}
\end{equation}
for $i,j \in \{+,-\}$. To classify the mutual visibility state between the two half-space radiating surfaces and determine whether the infinite plane containing one surface physically cuts through the finite boundaries of the other, we evaluate the signed distance from each Rx vertex to the Tx plane and from each Tx vertex to the Rx plane:
\begin{equation}
\begin{aligned}
d_{Rx}(i,j) &= (\mathbf{R}_{i,j} - \mathbf{T}^0) \cdot \hat{\mathbf{n}}_T \\
d_{Tx}(i,j) &= (\mathbf{T}_{i,j} - \mathbf{R}^0) \cdot \hat{\mathbf{n}}_R.
\end{aligned}
\end{equation}
If all $d_{Tx}(i,j) \le 0$ or all $d_{Rx}(i,j) \le 0$, then at least one surface lies entirely in the blind half-space or grazing plane of the other, yielding no visibility and zero DoF. In contrast, if all $d_{Tx}(i,j) > 0$ and all $d_{Rx}(i,j) > 0$, full visibility is achieved; since $\mathbf{T}^0$ and $\mathbf{R}^0$ are the averages of their respective four vertices, this condition automatically guaranties $\hat{\mathbf{n}}_T \cdot \mathbf{r}^0 > 0$ and $\hat{\mathbf{n}}_R \cdot (-\mathbf{r}^0) > 0$ (where $\mathbf{r}^0 = \mathbf{R}^0 - \mathbf{T}^0$), ensuring $|\det(\mathbf{A})| > 0$. In all other cases, partial visibility occurs, requiring any intersected surface to be clipped to obtain its effective area and shifted centroid. Once effective centers are established, the check $(\hat{\mathbf{n}}_T \cdot \mathbf{r}_{\text{eff}}) \le 0$ or $(\hat{\mathbf{n}}_R \cdot (-\mathbf{r}_{\text{eff}})) \le 0$ serves as a final numerical guard against grazing-incidence singularities ($\det(\mathbf{A}) = 0$). This is achieved by calculating the vertices of the convex polygons. A point on the Tx surface is visible to the Rx plane if it lies in the forward half-space of Rx, i.e., if $(\mathbf{T}(\eta_x, \eta_y) - R^0) \cdot \hat{\mathbf{n}}_R > 0$.
Substituting $\mathbf{T}(\eta_x, \eta_y) = \eta_x \hat{\mathbf{u}}_{Tx} + \eta_y \hat{\mathbf{u}}_{Ty}$ produces a linear boundary equation in the local Tx coordinates:
$$(\hat{\mathbf{u}}_{Tx} \cdot \hat{\mathbf{n}}_R)\eta_x + (\hat{\mathbf{u}}_{Ty} \cdot \hat{\mathbf{n}}_R)\eta_y = \mathbf{R}^0 \cdot \hat{\mathbf{n}}_R$$
Let $A_T = \hat{\mathbf{u}}_{Tx} \cdot \hat{\mathbf{n}}_R$, $B_T = \hat{\mathbf{u}}_{Ty} \cdot \hat{\mathbf{n}}_R$, and $C_T = \mathbf{R}^0 \cdot \hat{\mathbf{n}}_R$. The cutting line on the Tx plane is then $A_T \eta_x + B_T \eta_y = C_T$, and the visible region satisfies $A_T \eta_x + B_T \eta_y > C_T$.
Similarly, a point on the Rx surface is visible if it lies in the forward half-space of Tx, i.e., if $(\mathbf{R}(\zeta_x, \zeta_y) - \mathbf{T}^0) \cdot \hat{\mathbf{n}}_T > 0$. Substituting $\mathbf{R}(\zeta_x, \zeta_y) = \mathbf{R}^0 + \zeta_x \hat{\mathbf{u}}_{Rx} + \zeta_y \hat{\mathbf{u}}_{Ry}$ and $\mathbf{T}^0 = (0,0,0)$ yields:
$$(\hat{\mathbf{u}}_{Rx} \cdot \hat{\mathbf{n}}_T)\zeta_x + (\hat{\mathbf{u}}_{Ry} \cdot \hat{\mathbf{n}}_T)\zeta_y = -\mathbf{R}^0 \cdot \hat{\mathbf{n}}_T$$
Let $A_R = \hat{\mathbf{u}}_{Rx} \cdot \hat{\mathbf{n}}_T$, $B_R = \hat{\mathbf{u}}_{Ry} \cdot \hat{\mathbf{n}}_T$, and $C_R = -\mathbf{R}^0 \cdot \hat{\mathbf{n}}_T$. The cutting line on the Rx plane is then $A_R \zeta_x + B_R \zeta_y = C_R$, and the visible region satisfies $A_R \zeta_x + B_R \zeta_y > C_R$. To determine the vertices of the clipped polygon, we apply the cutting lines to the original rectangular boundaries. We will start with Tx as an example, and the same process applies to Rx. Initially, we evaluate the visibility condition $A_T \eta_x + B_T \eta_y > C_T$ for the four original corners, $(\pm \frac{L_{Tx}}{2}, \pm \frac{L_{Ty}}{2})$, and keep the corners that satisfy the condition. Then we calculate where the cutting line, $A_T \eta_x + B_T \eta_y = C_T$, intersects the four boundary edges of the rectangle. For the left/right edges $\left(\eta_x = \pm \frac{L_{Tx}}{2}\right)$, if $B_T \neq 0$, we obtain $\eta_y = \frac{C_T \mp A_T(L_{Tx}/2)}{B_T}$, whereas if $B_T = 0$, the cutting line is parallel to these edges and yields no intersection. Similarly, for the upper/lower edges $\left(\eta_y = \pm \frac{L_{Ty}}{2}\right)$, if $A_T \neq 0$, we obtain $\eta_x = \frac{C_T \mp B_T(L_{Ty}/2)}{A_T}$, whereas if $A_T = 0$, the cutting line is parallel to these edges and yields no intersection. Note that $A_T$ and $B_T$ cannot vanish simultaneously under partial visibility, as $A_T = B_T = 0$ implies parallel Tx and Rx planes that cannot intersect. We then keep only the valid intersection points falling within the physical limits of the rectangle, i.e., $\eta_y \in [-L_{Ty}/2, L_{Ty}/2]$ for the left/right edges and $\eta_x \in [-L_{Tx}/2, L_{Tx}/2]$ for the upper/lower edges, removing any duplicate vertices if the cutting line passes through an original corner. Finally, we combine the survived original corners and the valid intersection points of the edges and order these $N$ vertices counterclockwise to define the new effective convex polygon: $\{ (\eta_{x,1}, \eta_{y,1}), (\eta_{x,2}, \eta_{y,2}), \dots, (\eta_{x,N}, \eta_{y,N}) \}$. To close the loop, we set $(\eta_{x,N+1}, \eta_{y,N+1}) = (\eta_{x,1}, \eta_{y,1})$. Now, the effective area on the Tx is calculated by
\begin{equation}
	S_{T, eff} = \frac{1}{2} \sum_{i=1}^{N} (\eta_{x,i} \eta_{y,i+1} - \eta_{x,i+1} \eta_{y,i}).
	\end{equation}
The local coordinates of the Tx centroid are given by
\begin{equation}
	\begin{aligned}
		& \eta_{x,c} = \frac{1}{6 S_{T, eff}} \sum_{i=1}^{N} (\eta_{x,i} + \eta_{x,i+1})(\eta_{x,i} \eta_{y,i+1} - \eta_{x,i+1} \eta_{y,i}) \\
		& \eta_{y,c} = \frac{1}{6 S_{T, eff}} \sum_{i=1}^{N} (\eta_{y,i} + \eta_{y,i+1})(\eta_{x,i} \eta_{y,i+1} - \eta_{x,i+1} \eta_{y,i}).
	\end{aligned}
\end{equation}
The same procedure is followed on the Rx surface for the calculation of the $S_{R,eff}$ and the coordinates $\zeta_{x,c},\zeta_{y,c}$, with $M$ now being the number of ordered local vertices. Finally, we map the local 2D centroids back into the 3D space to establish the updated $\mathbf{T}^0_{eff}$ and $\mathbf{R}^0_{eff}$
\begin{equation}
	\begin{aligned}
		& \mathbf{T}^0_{eff} = \eta_{x,c} \hat{\mathbf{u}}_{Tx} + \eta_{y,c} \hat{\mathbf{u}}_{Ty} \\
		& \mathbf{R}^0_{eff} = \mathbf{R}^0 + \zeta_{x,c} \hat{\mathbf{u}}_{Rx} + \zeta_{y,c} \hat{\mathbf{u}}_{Ry}.
	\end{aligned}
\end{equation}
The calculation steps are shown in Algorithm 1.
\begin{algorithm}[t]
\caption{Check Visibility Status for DoF Calculation}
\label{alg:visibility}
\begin{algorithmic}[1]
\STATE Calculate $\mathbf{T}_{i,j}$, $\mathbf{R}_{i,j}$ via (49), the normal vectors $\hat{\mathbf{n}}_T$, $\hat{\mathbf{n}}_R$, and the signed distances $d_{Tx}(i,j)$, $d_{Rx}(i,j)$ via (50).
\IF{ALL $d_{Tx}(i,j) \le 0$ \textbf{or} ALL $d_{Rx}(i,j) \le 0$}
    \STATE $\text{DoF} \leftarrow 0$ \COMMENT{No visibility: blind half-space or grazing limit}
\ELSIF{ALL $d_{Tx}(i,j) > 0$ \textbf{and} ALL $d_{Rx}(i,j) > 0$}
    \STATE $\mathbf{r}_{eff} \leftarrow \mathbf{R}^0 - \mathbf{T}^0$, $d_{eff} \leftarrow \|\mathbf{r}_{eff}\|$
    \STATE Calculate $|\det(\mathbf{A})|$ using (30) and $\text{DoF}$ using (29) (or (48)) \COMMENT{Full visibility}
\ELSE[Partial visibility]
    \IF{ANY $d_{Tx}(i,j) \le 0$}
        \STATE Clip Tx rectangle keeping $(\mathbf{T}(\eta_x,\eta_y) - \mathbf{R}^0)\cdot\hat{\mathbf{n}}_R > 0$.
        \STATE Compute $S_{T,eff}$, $(\eta_{x,c},\eta_{y,c})$, and $\mathbf{T}_{eff}^0$ via (51)--(53).
    \ELSE
        \STATE $S_{T,\text{eff}} \leftarrow L_{Tx}L_{Ty}$, $\mathbf{T}_{eff}^0 \leftarrow \mathbf{T}^0$
    \ENDIF
    \IF{ANY $d_{Rx}(i,j) \le 0$}
        \STATE Clip Rx rectangle keeping $(\mathbf{R}(\zeta_x,\zeta_y) - \mathbf{T}^0)\cdot\hat{\mathbf{n}}_T > 0$.
        \STATE Compute $S_{R,eff}$, $(\zeta_{x,c},\zeta_{y,c})$, and $\mathbf{R}_{eff}^0$ via (51)--(53).
    \ELSE
        \STATE $S_{R,eff} \leftarrow L_{Rx}L_{Ry}$, $\mathbf{R}_{eff}^0 \leftarrow \mathbf{R}^0$
    \ENDIF
    \STATE $\mathbf{r}_{eff} \leftarrow \mathbf{R}_{eff}^0 - \mathbf{T}_{eff}^0$, $d_{\text{eff}} \leftarrow \|\mathbf{r}_{eff}\|$
    \IF{$(\hat{\mathbf{n}}_T \cdot \mathbf{r}_{eff}) \le 0$ \textbf{or} $(\hat{\mathbf{n}}_R \cdot (-\mathbf{r}_{eff})) \le 0$}
        \STATE $\text{DoF} \leftarrow 0$ \COMMENT{Grazing-incidence guard ($\det(\mathbf{A}) = 0$)}
    \ELSE
        \STATE Calculate $|\det(\mathbf{A})|$ using (30) and $\text{DoF}$ using (35) (or (48)).
    \ENDIF
\ENDIF
\end{algorithmic}
\end{algorithm}
\section{Conclusions}
In this paper, we presented a closed-form deterministic framework for evaluating scalar and vector spatial DoF of planar holographic MIMO systems under arbitrary 3D placements and orientations. Accounting for half-space radiation, we developed a geometric clipping algorithm that classifies full, partial, and zero visibility, extracts convex polygons, and computes effective areas and shifted centroids. Anchoring a double-sided quadratic Taylor distance expansion to these centroids mitigates phase truncation errors, beam defocusing, and DoF leakage in near-field and clipped regimes. From the spatial kernel, we derived a projection matrix $\mathbf{A}$ whose determinant shows how tilt skews spatial coherence cells into fractional DoF. Extending the model to dual-polarized vector fields via the radiative dyadic Green's function yielded a Kronecker-factored tensor kernel quantifying the polarization gain. Future work includes relaxing the uniform-amplitude distance constraint and embedding this model into stochastic geometry analyses.
\appendices
\section{Proof of Equation (\ref{kernel_2})} 
If we neglect the constant $1/\left(4 \pi d_{eff}\right)^2$, (\ref{kernel_1}) is written as
\begin{equation} \label{app_kernel_1}
	K_R(\tilde{\pmb{\zeta}},\tilde{\pmb{\zeta}}') \propto \iint_{S_{T,eff}} {\rm{exp}}\left[-jk \left( d(\tilde{\pmb{\eta}},\tilde{\pmb{\zeta}}) - d(\tilde{\pmb{\eta}},\tilde{\pmb{\zeta}}') \right)\right] \rm{d}\tilde{\eta}_x \rm{d}\tilde{\eta}_y
\end{equation}
We use the expansions of (\ref{Taylor}), (\ref{dlinear}), and (\ref{dquad}). The constant effective distance $d_{eff}$ is identical for both paths and cancels out. The same applies to the linear terms $(\tilde{\rho}_{Tx}\tilde{\eta}_x + \tilde{\rho}_{Ty}\tilde{\eta}_y)$ that cancel completely because they do not depend on $\tilde{\pmb{\zeta}}$ and $\tilde{\pmb{\zeta}}'$. The remaining components of the linear term are
\begin{equation}
	\begin{aligned}
	d_{linear}(\tilde{\pmb{\eta}},\tilde{\pmb{\zeta}}) - d_{linear}(\tilde{\pmb{\eta}},\tilde{\pmb{\zeta}}') & = \tilde{\rho}_{Rx}(\tilde{\zeta}_x -\tilde{\zeta}'_x) + \tilde{\rho}_{Ry}(\tilde{\zeta}_y-\tilde{\zeta}'_y) \\ & = \tilde{\rho}_{Rx} \Delta\tilde{\zeta}_x + \tilde{\rho}_{Ry} \Delta\tilde{\zeta}_y
	\end{aligned}
\end{equation}
This term is independent of the integration variables $\tilde{\eta}_x, \tilde{\eta}_y$ and factored out of the integral as a global scalar phase $C_{linear}$. Regarding the quadratic term, the first part cancels out since it is independent of $\tilde{\pmb{\zeta}}$ and $\tilde{\pmb{\zeta}}'$. The second part is given by
\begin{equation}
	\begin{aligned}
	  \frac{1}{2d_{eff}}\Big\{ & \left(\tilde{\zeta}_x^2 - \tilde{\zeta'}_x^2\right)\left(1 - \tilde{\rho}_{Rx}^2\right) + \left(\tilde{\zeta}_y^2 - \tilde{\zeta'}_y^2\right)\left(1 - \tilde{\rho}_{Ry}^2\right) -  \Biggr. \\  & \Biggl. 2\tilde{\rho}_{Rx}\tilde{\rho}_{Ry} \left(\tilde{\zeta}_x\tilde{\zeta}_y - \tilde{\zeta'}_x\tilde{\zeta'}_y\right)\Bigr\}
	 \end{aligned}
\end{equation}
which is independent of the integration variables $\tilde{\eta}_x, \tilde{\eta}_y$ and factored out of the integral as a global scalar phase $C_{quad}$.
The third term is a cross-coupling term that is written as
\begin{equation}
\begin{aligned}
\frac{-1}{d_{eff}} \sum_{i \in \{x,y\}} \sum_{j \in \{x,y\}} & \tilde{\eta}_i (\tilde{\zeta}_j - \tilde{\zeta}'_j) \left( \hat{\mathbf{u}}_{Ti} \cdot \hat{\mathbf{u}}_{Rj} - \tilde{\rho}_{Ti}\tilde{\rho}_{Rj} \right) \\ 
& = -\tilde{\eta}_x \frac{\tilde{K}_{\rho,x}}{k} - \tilde{\eta}_y \frac{\tilde{K}_{\rho,y}}{k}
\end{aligned}
\end{equation}
where $\tilde{K}_{\rho,x}$ and $\tilde{K}_{\rho,y}$ are given in (\ref{Krhox}) and (\ref{Krhoy}) respectively. Therefore, the kernel simplifies to
\begin{equation}
	\begin{aligned}
K_R(\Delta\tilde{\zeta}_x,\Delta\tilde{\zeta}_y) & \propto e^{-jk\left(C_{linear}+C_{quad} \right)} \\ & \iint_{S_{T,eff}} e^{j (\tilde{K}_{\rho,x}\tilde{\eta}_x + \tilde{K}_{\rho,y}\tilde{\eta}_y)} d\tilde{\eta}_x d\tilde{\eta}_y 
	\end{aligned}
\end{equation}
and since $\vert{}\exp(-jk(C_{linear} + C_{quad}))\vert{} = 1$ then (\ref{kernel_2}) results.
\section{Proof of Equation (\ref{detA})}
The projection matrix $\mathbf{A}$ is written as the product of the local bases $\mathbf{U}_T = [\hat{\mathbf{u}}_{Tx},\hat{\mathbf{u}}_{Ty}]$, $\mathbf{U}_R = [\hat{\mathbf{u}}_{Rx},\hat{\mathbf{u}}_{Ry}]$, and a projection operator that projects any vector onto the plane orthogonal to the propagation path, i.e., $\mathbf{P} = \mathbf{I}-\hat{\mathbf{r}}_{eff}\hat{\mathbf{r}}_{eff}^T$,
\begin{equation}\label{matrixA}
\mathbf{A} = \mathbf{U}_T^T \mathbf{P} \mathbf{U}_R
\end{equation}
Indeed, 
\begin{equation}
	\begin{aligned}
	A_{ij} & = \hat{\mathbf{u}}_{Ti} \cdot \hat{\mathbf{u}}_{Rj} - \tilde{\rho}_{Ti}\tilde{\rho}_{Rj} \\
	& = \hat{\mathbf{u}}_{Ti}^T \mathbf{I} \hat{\mathbf{u}}_{Rj} -\left( \hat{\mathbf{u}}_{Ti}^T \hat{\mathbf{r}}_{eff} \right) \left( \hat{\mathbf{r}}_{eff}^T \hat{\mathbf{u}}_{Rj} \right) \\
	& = \hat{\mathbf{u}}_{Ti}^T \left( \mathbf{I} - \hat{\mathbf{r}}_{eff}\hat{\mathbf{r}}_{eff}^T \right) \hat{\mathbf{u}}_{Rj}
	 = \hat{\mathbf{u}}_{Ti}^T \mathbf{P} \hat{\mathbf{u}}_{Rj}
	\end{aligned}
\end{equation}
Therefore, $A_{ij}$ is the dot product of both axes after they have both been projected onto the orthogonal plane.
Geometrically, the channel ignores the true 3D angle between the Tx and Rx antennas; it only considers how well their projections align on the 2D plane perpendicular to the wave. If the Tx shadow and the Rx shadow align perfectly in that plane, $A_{ij}$ is maximized. Because $\mathbf{P}$ projects any vector onto the 2D plane perpendicular to $\hat{\mathbf{r}}_{eff}$, we can construct a completely generic, arbitrary orthonormal basis for this plane $\mathbf{E}=[\hat{\mathbf{e}}_1,\hat{\mathbf{e}}_2]$. The two unit vectors  $\hat{\mathbf{e}}_1$ and $\hat{\mathbf{e}}_2$ are orthogonal to each other, orthogonal to $\hat{\mathbf{r}}_{eff}$, and $\hat{\mathbf{e}}_1 \times \hat{\mathbf{e}}_2 = \hat{\mathbf{r}}_{eff}$. Then, we can use the exact projector factorization $\mathbf{P}=\mathbf{EE}^T$ and (\ref{matrixA}) can be written as
\begin{equation}
\begin{aligned}
	& \mathbf{A} = \mathbf{U}_T^T \mathbf{P} \mathbf{U}_R=\left(\mathbf{U}_T^T \mathbf{E}\right) \left(\mathbf{E}^T\mathbf{U}_R\right) \\
	& \Rightarrow {\rm{det}}(\mathbf{A})={\rm{det}}\left(\mathbf{U}_T^T \mathbf{E}\right) {\rm{det}}\left(\mathbf{E}^T\mathbf{U}_R\right)
\end{aligned}
\end{equation}
Therefore,
\begin{equation}
	\left(\mathbf{U}_T^T \mathbf{E}\right)=\begin{bmatrix} \hat{\mathbf{u}}_{Tx}^T \\ \hat{\mathbf{u}}_{Ty}^T \end{bmatrix} \begin{bmatrix} \hat{\mathbf{e}}_1 & \hat{\mathbf{e}}_2 \end{bmatrix} = \begin{bmatrix} \hat{\mathbf{u}}_{Tx} \cdot \hat{\mathbf{e}}_1 & \hat{\mathbf{u}}_{Tx} \cdot \hat{\mathbf{e}}_2 \\ \hat{\mathbf{u}}_{Ty} \cdot \hat{\mathbf{e}}_1 & \hat{\mathbf{u}}_{Ty} \cdot \hat{\mathbf{e}}_2 \end{bmatrix}
\end{equation}
and the determinant is given by
\begin{equation}
	\det\left(\mathbf{U}_T^T \mathbf{E}\right) = (\hat{\mathbf{u}}_{Tx} \cdot \hat{\mathbf{e}}_1)(\hat{\mathbf{u}}_{Ty} \cdot \hat{\mathbf{e}}_2) - (\hat{\mathbf{u}}_{Tx} \cdot \hat{\mathbf{e}}_2)(\hat{\mathbf{u}}_{Ty} \cdot \hat{\mathbf{e}}_1).
\end{equation}
Using Lagrange's vector identity
\begin{equation}
	(\mathbf{a} \cdot \mathbf{c})(\mathbf{b} \cdot \mathbf{d}) - (\mathbf{a} \cdot \mathbf{d})(\mathbf{b} \cdot \mathbf{c}) = (\mathbf{a} \times \mathbf{b}) \cdot (\mathbf{c} \times \mathbf{d})
\end{equation}
the determinant is given by
\begin{equation}
	\begin{aligned}
	{\rm{det}}\left(\mathbf{U}_T^T \mathbf{E}\right) &= (\hat{\mathbf{u}}_{Tx} \times \hat{\mathbf{u}}_{Ty}) \cdot (\hat{\mathbf{e}}_1 \times \hat{\mathbf{e}}_2) \\
	&= \hat{\mathbf{n}}_T \cdot \hat{\mathbf{r}}_{eff} = \cos(\gamma_T)
		\end{aligned}
\end{equation}
Using similar steps
\begin{equation}
	\begin{aligned}
		{\rm{det}}\left(\mathbf{E}^T\mathbf{U}_R\right)& = (\hat{\mathbf{e}}_1 \times \hat{\mathbf{e}}_2) \cdot (\hat{\mathbf{u}}_{Rx} \times \hat{\mathbf{u}}_{Ry}) \\
		&= \hat{\mathbf{r}}_{eff} \cdot \hat{\mathbf{n}}_R = -\cos(\gamma_R)
	\end{aligned}
\end{equation}
where $\gamma_R$ is measured from $-\hat{\mathbf{r}}_{eff}$. Finally, (\ref{detA}) yields.
\section{Proof of Equation (\ref{partialkernel})}
We need to evaluate the surface integral of the phase over an arbitrary polygonal transmitting surface $S_{T,eff}$ with $N$ edges
\begin{equation}
I(\tilde{\mathbf{K}}_\rho) = \iint_{S_{T,eff}} e^{j \tilde{\mathbf{K}}_\rho \cdot \boldsymbol{\eta}} \, {\rm{d}}A
\end{equation}
To convert this surface integral into a line integral along the polygon's edges, we use the Green's theorem:
\begin{equation}\label{Green'stheo}
	\iint_{S_{T,eff}} (\nabla \cdot \mathbf{F}) \, {\rm{d}}A = \oint_{P} (\mathbf{F} \cdot \hat{\mathbf{n}}) \, {\rm{d}}l
\end{equation}
where the vector field $\mathbf{F}$ is set to:
\begin{equation}
\mathbf{F} = \frac{\tilde{\mathbf{K}}_\rho}{j \vert{}\tilde{\mathbf{K}}_\rho\vert{}^2} e^{j \tilde{\mathbf{K}}_\rho \cdot \boldsymbol{\eta}}	
\end{equation}
Then we substitute $\mathbf{F}$ into (\ref{Green'stheo}):
\begin{equation}
I(\tilde{\mathbf{K}}_\rho) = \oint_{P} \left( \frac{\tilde{\mathbf{K}}_\rho}{j \vert{}\tilde{\mathbf{K}}_\rho\vert{}^2} e^{j \tilde{\mathbf{K}}_\rho \cdot \boldsymbol{\eta}} \right) \cdot \hat{\mathbf{n}} \, {\rm{d}}l	
\end{equation}
Since the boundary consists of $N$ straight edges, we break the loop into a summation over each edge $i$:
\begin{equation}\label{lineintegral}
	I(\tilde{\mathbf{K}}_\rho) = \frac{1}{j \vert{}\tilde{\mathbf{K}}_\rho\vert{}^2} \sum_{i=1}^N (\hat{\mathbf{n}}_{P,i} \cdot \tilde{\mathbf{K}}_\rho) \int_{L_i} e^{j \tilde{\mathbf{K}}_\rho \cdot \boldsymbol{\eta}} \, {\rm{d}}l
\end{equation}
where $\hat{\mathbf{n}}_{P,i}$ is the unit outward normal vector for edge $i$, which is constant along that straight edge, so it pulls out of the integral. Let edge $i$ be defined by its midpoint $\tilde{\mathbf{C}}_i$ and its full vector length $\Delta\mathbf{V}_i$. Any point $\boldsymbol{\eta}$ on this edge can be parameterized by a scalar $t$ from $-1/2$ to $1/2$:
$$\boldsymbol{\eta}(t) = \tilde{\mathbf{C}}_i + t \Delta\mathbf{V}_i$$
The differential arc length is ${\rm{d}}l = \vert{}\Delta\mathbf{V}_i\vert{} {\rm{d}}t$. Substitute this into the line integral:
\begin{equation}
	\begin{aligned}
\int_{L_i} e^{j \tilde{\mathbf{K}}_\rho \cdot \boldsymbol{\eta}} \, {\rm{d}}l & = \int_{-1/2}^{1/2} e^{j \tilde{\mathbf{K}}_\rho \cdot (\tilde{\mathbf{C}}_i + t \Delta\mathbf{V}_i)} \vert{}\Delta\mathbf{V}_i\vert{} \, {\rm{d}}t \\ &= \vert{}\Delta\mathbf{V}_i\vert{} e^{j \tilde{\mathbf{K}}_\rho \cdot \tilde{\mathbf{C}}_i} \int_{-1/2}^{1/2} e^{j (\tilde{\mathbf{K}}_\rho \cdot \Delta\mathbf{V}_i) t} \, {\rm{d}}t	\\
&= \vert{}\Delta\mathbf{V}_i\vert{} e^{j \tilde{\mathbf{K}}_\rho \cdot \tilde{\mathbf{C}}_i} \text{sinc}\left( \frac{\tilde{\mathbf{K}}_\rho \cdot \Delta\mathbf{V}_i}{2\pi} \right)
	\end{aligned}
\end{equation}
Substituting in (\ref{lineintegral}) yields (\ref{partialkernel}). 
\section*{Acknowledgment}
This work was supported by the European Commission through the Horizon Europe project QUEST-6G under the GA No. 101292676, and the University of Piraeus Research Center. Parts of this manuscript were drafted with the help of an AI language model (Google Gemini); the author reviewed, verified, and takes full responsibility for all content.
\end{document}